\RequirePackage{ifpdf}
\documentclass{JHEP3}

\usepackage{arydshln}

\usepackage{amsfonts,amssymb,amsmath,bm}
\usepackage{slashed}
\usepackage{graphicx}

\ifpdf
{}
\else
\fi

\usepackage{epstopdf}

\newcommand{\beq}{\begin{equation}}
\newcommand{\eeq}{\end{equation}}
\newcommand{\beqq}{\begin{equation*}}
\newcommand{\eeqq}{\end{equation*}}
\newcommand\beqa{\begin{eqnarray}}
\newcommand\eeqa{\end{eqnarray}}
\newcommand\beqaa{\begin{eqnarray*}}
\newcommand\eeqaa{\end{eqnarray*}}
\newcommand\bea{\begin{array}}
\newcommand\eea{\end{array}}
\newcommand\beaa{\begin{array}}
\newcommand\eeaa{\end{array}}

\def\XXint#1#2#3{{\setbox0=\hbox{$#1{#2#3}{\int}$ }
\vcenter{\hbox{$#2#3$ }}\kern-.5\wd0}}

\def\XXint#1#2#3{{\setbox0=\hbox{$#1{#2#3}{\int}$}
\vcenter{\hbox{$#2#3$}}\kern-.5\wd0}}

\newcommand{\nn}{\nonumber}

\newcommand{\neqa}{\nonumber\end{eqnarray}}
\newcommand{\la}[1]{\label{#1}}

\newcommand{\eq}[1]{(\ref{#1})}

\newcommand{\hs}{\frac{\sqrt{3}}{2}}
\renewcommand{\d}{\partial}

\newcommand{\<}{{\langle}}
\renewcommand{\>}{{\rangle}}

\newcommand{\re}{\relax{\rm I\kern-.18em R}}

\renewcommand{\sp}{p\hspace{-.40em}/}

\def\su2{{SU(2)}}

\def\eps{{\epsilon}}

\def\[{\left[}
\def\]{\right]}

\def\s{\sigma}

\def\({\left(}
\def\){\right)}
\def\[{\left[}
\def\]{\right]}

\def\<{\langle}
\def\>{\rangle}

\def\cO{{\cal O}}

\def\s*{\ *_{\!\!\!\!\!\!\!\!\!\,_{\,_\text{\scriptsize{sym}}}}}
\def\hs*{\ \hat{*}_{\!\!\!\!\!\!\!\!\!\,_{\,_\text{\scriptsize{sym}}}}}
\def\d{\partial}

\def\i2{\frac{i}{2}}

\def\bQ{{\bf Q}}
\def\bP{{\bf P}}

\def\bq{{\bf q}}
\def\bp{{\bf p}}

\def\spi{\relax{\rm \pi\kern-0.5em /}}
\def\sA{\relax{\rm A\kern-0.5em /}}
\def\sp{\relax{\rm p\kern-0.5em /}}
\def\sd{\relax{\rm \d\kern-0.5em /}}
\def\sk{\relax{\rm k\kern-0.5em /}}
\def\sn{\relax{\rm n\kern-0.5em /}}
\def\sl{\relax{\rm l\kern-0.5em /}}
\def\sP{\relax{\rm P\kern-0.7em /}}
\def\sBethe{\relax{\rm \Bethe\kern-0.5em /}}
\def\cN{{\cal N}}

    \def\const{\mbox{const}}

	\newcommand{\mZ}{{\mathbb Z}}

	\renewcommand{\Re}{{\rm Re}}

\def\d{\partial}

\newcommand{\Gc}{{\Gamma_\text{cusp}}}

\title{
Quantum Spectral Curve for a Cusped Wilson Line in ${\mathcal N}=4$ SYM
}

\author{Nikolay Gromov$^{1,2}$,
  Fedor Levkovich-Maslyuk$^{1}$ \\
  $^1$King's College London, Department of Mathematics, \\ The Strand, London WC2R 2LS,
  United Kingdom\\ \\
  $^2$ St.Petersburg INP, Gatchina, 188 300, St.Petersburg, Russia
  \qquad\\ \\
  \textit{E-mail:}
  \email{nikgromov$\bullet$gmail.com}, \email{fedor.levkovich$\bullet$gmail.com}\\ \\}

\abstract{
We show that the Quantum Spectral Curve (QSC) formalism, initially formulated for the spectrum of anomalous dimensions
of all local single trace operators in ${\cal N}=4$ SYM, can be extended to the generalized cusp anomalous dimension for all values of the parameters.
We find that the large spectral parameter asymptotics and some analyticity properties have to be modified, but the functional relations are unchanged.
As a demonstration, we find an all-loop analytic expression
for the first two nontrivial terms in the small $|\phi\pm \theta|$ expansion. We also present nonperturbative numerical results at generic angles which match perfectly $4$-loop perturbation theory and the classical string prediction. 

The reformulation of the problem in terms of the QSC opens the possibility to explore many open questions. We attach to this paper several Mathematica notebooks which should facilitate future studies.

}

\preprint{}

\begin{document}

\newpage

\section{Introduction}

The exploration of integrable structures in planar $\cN=4$ supersymmetric Yang-Mills theory has led to numerous results which go far beyond the usual restrictions of perturbative QFT calculations \cite{Beisert:2010jr}. 
With the help of integrability the spectral problem
was   reduced to a strikingly simple set of Riemann-Hilbert type equations known as the Quantum Spectral Curve (QSC) \cite{PmuPRL,PmuLong}. They are expected to capture the exact spectrum of single trace operator scaling dimensions and string state energies in the dual $AdS_5\times S^5$ theory at any value of the 't Hooft coupling $\lambda$. As a compact set of equations for only a few functions, the QSC is tremendously more efficient than the
preceding infinite system of Thermodynamic Bethe ansatz (TBA) or Y-system equations \cite{Gromov:2009tv,Bombardelli:2009ns,Gromov:2009bc,Arutyunov:2009ur,Cavaglia:2010nm}.

The efficiency of the Quantum Spectral Curve over any other approach has been already demonstrated in a variety of settings. In \cite{Marboe:2014gma,Marboe:2014sya} it was used to reach up to 10 loops in perturbation theory, while the all-loop near-BPS solution in \cite{Gromov:2014bva} led to new strong coupling predictions. At finite coupling it allows to compute the spectrum numerically with extremely high precision \cite{Gromov:2015wca}. In addition, the QSC made it finally possible to deeply probe the BFKL regime using integrability. The leading order BFKL predictions were reproduced in \cite{Alfimov:2014bwa} and very recently the novel NNLO term was computed in \cite{Gromov:2015vua}. The QSC has been formulated for the ABJM theory as well \cite{Cavaglia:2014exa}, leading to weak coupling results \cite{Anselmetti:2015mda} and to an exact computation \cite{Gromov:2014eha} of the interpolating function which enters all
integrability-based predictions.
Finally, originating in the universal QQ-relations, the QSC is also expected to be helpful in application to 3-point functions, as it should provide the exact wavefunctions in Sklyanin's separated variables.

A natural goal is to extend the Quantum Spectral Curve to various integrable deformations and boundary problems in $\cN=4$ SYM, making possible an in-depth investigation of their properties.
The TBA equations/Y-systems for these examples \cite{betadef,Arutyunov:2012zt,Arutyunov:2012ai,Arutynov:2014ota,Bajnok:2013wsa,Hegedus:2015kga,Bajnok:2013sza,Bajnok:2012xc,Zoubos:2010kh} are quite similar to those in the undeformed theory,
suggesting that the  QSC could also be adapted with relatively small changes.

In this paper we focus on one of the most intriguing problems of this kind, namely the boundary TBA for the generalized cusp anomalous dimension $\Gc$. This much-studied observable corresponds to the divergence associated with a pair of Wilson lines forming a cusp of arbitrary angle $\phi$,
\beq
	\left\langle W\right\rangle\sim\left(\frac{\Lambda_{IR}}{\Lambda_{UV}}\right)^{\Gamma_\text{cusp}},
\eeq
where $\Lambda_{IR,UV}$ are the infrared and ultraviolet cutoffs. The Wilson lines include an additional coupling to the six scalars of $\cN=4$ SYM, defined by two constant unit vectors $\vec n,\vec n_\theta\in{\mathbb R}^6$ corresponding to the two lines, with angle $\theta$ between these vectors (see Fig. \ref{fig:WLcusp}). We can write the cusped Wilson line explicitly as
\beq
W={\rm P}\exp\!\int\limits_{-\infty}^0\! dt\[i  A\cdot\dot{x}_q+\vec\Phi\cdot\vec n\,|\dot x_q|\]\times {\rm P}\exp\!\int\limits_0^\infty\!dt\[i A\cdot\dot x_{\bar q}+\vec\Phi\cdot\vec n_{\theta}\,|\dot x_{\bar q}|\],
\eeq
where $\vec \Phi$ is a vector made out of the six scalars $\Phi_i$, and $x_q(t)$, $x_{\bar q}(t)$ are straight lines which form an angle $\phi$ at the cusp.
\FIGURE[ht]
{\label{fig:WLcusp}

    \begin{tabular}{cc}
    \includegraphics[scale=0.3]{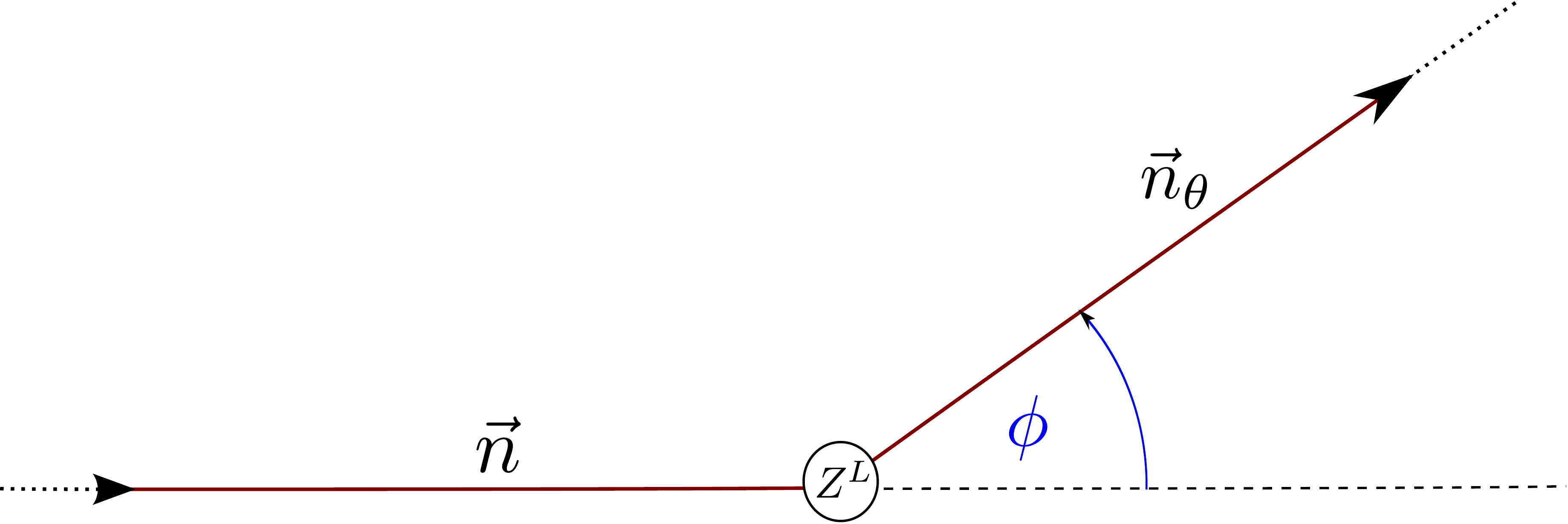}\\
    \end{tabular}
    \caption{\textbf{The cusped Wilson line.} A Wilson line with an angle $\phi$ at the cusp, with an extra insertion of $L$ scalar fields $Z=\Phi_1+i \Phi_2$. The coupling of the scalar fields to the two lines is determined by unit vectors $\vec n$ and $\vec n_\theta$ in the internal space, the angle between them is $\theta$.}
    }
The generalized cusp anomalous dimension can be equivalently understood as the quark-antiquark potential on the three-sphere and is related to many physical quantities, such as radiation power from a moving quark or IR divergences in amplitudes. Recently it was also found to determine the energy levels of a supersymmetric ``hydrogen atom'' made out of massive W-bosons in $\cN=4$ SYM \cite{Caron-Huot:2014gia}.

The key insight which allowed to derive exact equations for $\Gc$ from integrability was to consider the same Wilson lines with $L$ scalars $Z=\Phi_1+i\Phi_2$ inserted at the cusp\footnote{the scalars inserted at the cusp should be orthogonal to the combinations $\vec n\cdot\vec\Phi$ and $\vec n_\theta\cdot\vec\Phi$ which couple to the Wilson lines},
\beq
	\label{WilsL}
	W_L={\rm P}\exp\!\int\limits_{-\infty}^0\! dt\(i  A\cdot\dot{x}_q+\vec\Phi\cdot\vec n\,|\dot x_q|\)\times Z^L\times {\rm P}\exp\!\int\limits_0^\infty\!dt\(i A\cdot\dot x_{\bar q}+\vec\Phi\cdot\vec n_\theta\,|\dot x_{\bar q}|\).
\eeq
 This leads to a boundary problem for $\Gc$  as the ground state energy in finite volume $L$ with extra reflection phase factors corresponding to the two Wilson lines. The outcome is a set of TBA equations obtained in \cite{Correa:2012hh,Drukker:2012de} providing the value of $\Gc$ at any value of the coupling $\lambda$ for arbitrary $\phi,\theta$ and any number of insertions $L$. The usual definition of $\Gc$ corresponds to $L=0$.
 One can also consider more general insertions instead of $Z^L$ in \eq{WilsL}, which are described as excited states in the TBA.

When $\phi=\theta$ this observable is BPS, and one can study the near-BPS expansion in $(\phi-\theta)$ which can be written as
\beq
	\Gc=\frac{\cos\phi-\cos\theta}{\sin\phi}\Delta^{(1)}(\phi)
	+\(\frac{\cos\phi-\cos\theta}{\sin\phi}\)^2\Delta^{(2)}(\phi)+\cO((\phi-\theta)^3)
	\ .
\eeq

The leading coefficient in this series is known for $L=0$ to all loops from localization \cite{Correa:2012at,Fiol:2012sg}. For $\theta=0$ it was reproduced at any coupling in \cite{Gromov:2012eu} by a direct analytic solution of the TBA, which also gave a new prediction for the case with arbitrary $L$. This analytic solution was extended to the case with arbitrary $\theta\sim\phi$ and any $L$ in \cite{Gromov:2013qga}. The near-BPS results for $L\geq 1$ organize in a curious matrix model type partition function whose classical limit was investigated in \cite{Gromov:2012eu,Sizov:2013joa} giving the corresponding classical spectral curve (see also \cite{Beccaria:2013lca,Dekel:2013dy,Janik:2012ws}). In addition, the TBA was solved to two loops at weak coupling for finite $\phi,\theta$ \cite{Bajnok:2013sya}, reproducing direct perturbative predictions which are also known at up to four loops \cite{Makeenko:2006ds,Drukker:2011za,Correa:2012nk,Henn:2013wfa}.

In this paper we adapt the Quantum Spectral Curve approach to study the generalized cusp anomalous dimension at any values of the parameters.
Instead of deriving the QSC from the TBA, we make a proposal based on available data and consistency of the equations, and confirm it by several highly nontrivial tests. We find that all functional equations of the QSC remain unchanged, but the asymptotics at large values of the spectral parameter, as well as some of the analyticity properties, should be modified. In particular some functions acquire exponential asymptotics $\sim e^{\pm\phi u},e^{\pm\theta u}$,
as expected by analogy with spin chain Q-functions in the presence of twisted boundary conditions.
We also observed that rather subtle cancelations take place resulting
in  complicated constraints on subleading coefficients in the large $u$ asymptotics of $Q$-functions.
As an application we compute the subleading term (of order $(\phi-\theta)^2$) in the near-BPS expansion of $\Gc$ without scalar insertions, at any coupling and for any $\phi$. Our explicit result \eq{D2res} fully agrees with perturbative predictions.

Our paper is organized as follows. In section 2 we review the original Quantum Spectral Curve and discuss in detail the modifications needed for our problem. We discuss the vacuum state, i.e. with only $Z$ fields inserted at the cusp, but keep $L$ arbitrary. In section 3 we reconstruct the near-BPS solution at any $\theta$ and $L$, and then for $L=0$ extend it to the next order in the near-BPS expansion. In section 4 we describe a highly precise numerical method for solving the QSC equations and demonstrate it on several examples. Section 5 contains a discussion of the weak coupling solution at generic angles. We present conclusions in section 6. Several appendices contain various technical details, and we also attach to this paper several Mathematica notebooks.

\section{The Quantum Spectral Curve}

In this section we first briefly review the $\bP\mu$-subsystem of the original Quantum Spectral Curve equations for the spectrum of local operators in $\cN=4$ SYM (full details can be found in \cite{PmuLong}). Then we discuss how these equations should be modified to describe the generalised cusp anomalous dimension. We will see that all functional equations and analyticity conditions remain the same, and the only difference is in the asymptotics at large values of the spectral parameter.

The QSC is a system of functional equations for the exact Q-functions of the theory, which are analogs of the Baxter polynomials in spin chains. The analytic properties of these functions play a key role in the construction. A particularly simple basis of Q-functions is given by 4+4 functions $\bP_a(u)$ and $\bP^a(u)$ ($a=1,2,3,4$). Their very nice feature is that they have only one branch cut in the complex plane of the spectral parameter $u$. This is a `short' cut connecting the branch points at $u=\pm 2g$ where $g=\frac{\sqrt{\lambda}}{4\pi}$ is related to the 't Hooft coupling $\lambda$. Knowing $\bP_a(u)$ and $\bP^a(u)$ one can reconstruct all other Q-functions, which are analytic in the upper half-plane, but typically have infinitely many branch cuts in the lower half-plane.

\FIGURE[ht]{
\label{fig:pcuts}
\includegraphics[scale=0.18]{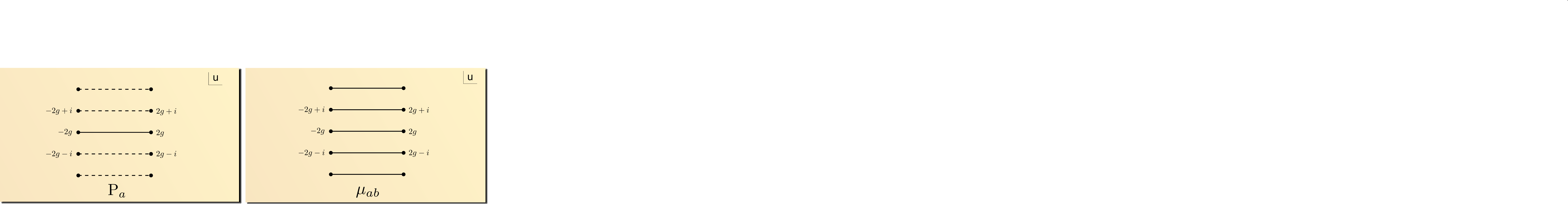}
\caption{ The branch cuts of $\bP_a$ and $\mu_{ab}$. While $\bP_a$ have only one cut, $\tilde\bP_a$ have infinitely many cuts, shown by dashed lines.}
}

While $\bP_a$ themselves have only one branch cut, by analytically continuing them through the cut we get new functions denoted as $\tilde\bP_a$, which now have infinitely many cuts at $u\in[-2g+in,2g+in],\ n\in\mZ$ (see Fig. \ref{fig:pcuts}). Remarkably, one can get a closed system of equations by introducing an antisymmetric matrix $\mu_{ab}(u)$ with unit Pfaffian, which relates $\tilde\bP_a$ to the original $\bP$-functions:
\beq
\label{tPd}
	\tilde\bP_a=\mu_{ab}\bP^b\;.
\eeq
The discontinuity of $\mu_{ab}$ on the cut is again expressed in terms of $\bP_a$ and $\mu_{ab}$,
\beq
\label{tmud}
	\tilde\mu_{ab}-\mu_{ab}=\tilde\bP_a\bP_b-\bP_a\tilde\bP_b\ .
\eeq
The functions $\mu_{ab}(u)$ have infinitely many cuts, but are $i$-periodic if the cuts are chosen to be `long' (connecting the branch points through infinity). With short cuts we have instead $\tilde\mu_{ab}(u)=\mu_{ab}(u+i)$. There are also similar equations for $\bP$-functions with upper indices,
\beq
\label{tup}
	\tilde\bP^a=\mu^{ab}\bP_b,\ \ \ \
	\tilde\mu^{ab}-\mu^{ab}=\tilde\bP^a\bP^b-\bP^a\tilde\bP^b
\eeq
where $\mu^{ab}$ is the inverse matrix, $\mu^{ab}\mu_{bc}=\delta^a_c$. Apart from the branch points, the $\bP$- and $\mu$-functions have no singularities in the complex plane. Finally, $\bP$'s have to satisfy
\beq
\label{Papa}
	\bP_a\bP^a=0\ .
\eeq
For left-right symmetric states (e.g. in the $sl(2)$ sector) the $\bP$'s with lower and upper indices are not independent:
\beq
\label{PcP}
	\bP^{a}=\chi^{ab}\bP_b
\eeq
where
\beq
\label{defchi}
	\chi^{ab}=\begin{pmatrix}
	0&0&0&-1\\
	0&0&1&0\\
	0&-1&0&0\\
	1&0&0&0
	\end{pmatrix}\ .
\eeq

The relations \eq{tPd}, \eq{tmud}, \eq{tup}, \eq{Papa} are known as the $\bP\mu$ system. In the $sl(2)$ sector they can be derived from the TBA equations, and in general stand as a conjecture which has been confirmed by
numerous tests.  It is important that the $\bP\mu$ system is a closed set of equations once it is supplemented by asymptotic constraints at $u\to\infty$ which in particular allow to compute the conformal dimension. These constraints are discussed in detail in the next section.

\subsection{Asymptotics in the original QSC for local operators}
To ensure uniqueness of the solution one should supplement the functional equations described above with asymptotic boundary conditions at $u\to\infty$. The asymptotics encode the Noether charges of the state, including the conformal dimension $\Delta$ (in our problem the analog of $\Delta$ is $\Gc$). Let us recall the derivation of the asymptotic constraints \cite{PmuLong} in some detail, as this step is a crucial point in adapting the QSC to the generalized cusp.

First, all Q-functions as well as $\mu_{ab}$ have power-like behavior at infinity
, and the powers in the asymptotics of $\bP$-functions encode the $SO(6)$ R-charges (or the angular momenta of the string on the $S^5$ part of the background)
\beq
\label{asymptP}
	\bP_a\sim A_au^{-\tilde M_a}, \ \bP^a\sim A^au^{\tilde M_a-1}
\eeq
with
\begin{align}
\label{Mta}
\tilde M_a=
&\left\{\frac{J_1+J_2-J_3+2}{2}
  ,\frac{J_1-J_2+J_3}{2}
   ,\frac{-J_1+J_2+J_3+2}{2}
   ,\frac{-J_1-J_2-J_3}{2}
   \right\}\ .
\end{align}
The charge $J_1$ also defines the length $L=J_1$ of the weak-coupling spin chain.

To constrain the leading coefficients $A_a,A^a$ it is very useful to consider another four Q-functions, denoted as $\bQ_i$, which are analogous to $\bP_a$ but correspond to dynamics in $AdS_5$ instead of $S^5$. Like $\bP_a$, these functions can be chosen to have only a single cut, but then it has to be a long cut at $u\in(-\infty,-2g]\cup[2g,+\infty)$. The $\bQ_i$ can be reconstructed from $\bP^a,\bP_a$ as
\beq
\label{QQP}
	\bQ_i=-\bP^aQ_{a|i}
\eeq
with the functions $Q_{a|i}$ obtained by solving the difference equation
\beq\la{Qai}
	{ Q}_{a|i}^+-{ Q}_{a|i}^-=-
	\bP_a
	\bP^b
	{ Q}_{b|i}^+\ \
\eeq
where we introduced the notation
\beq
	f^\pm=f(u\pm i/2),\ \ f^{[a]}=f(u+ia/2)\;.
\eeq
This is a 4th order difference equation which mixes the $Q_{a|i}$ corresponding to different values of $a$. The index $i$ then labels the 4 solutions of this equation. Using \eq{QQP} we can also rewrite
it as
\beq
\label{QPQ}
	{ Q}_{a|i}^+-{ Q}_{a|i}^-=\bP_a\bQ_i
\eeq
which is in fact one of the canonical QQ-relations.  The matrix $Q_{a|i}$ should be normalized such that it preserves the $\chi^{ab}$ matrix from \eq{defchi},
\beq
\label{cqcq}
	\chi Q \chi Q^T=1
\eeq
and should have unit determinant \cite{PmuLong},
\beq
\label{detQ1}
	\det_{1\leq a,i \leq 4}Q_{a|i}=1\ .
\eeq

Similarly to \eq{asymptP}, the powers in the asymptotics of $\bQ_i$ contain the $SO(4,2)$ charges including the conformal dimension $\Delta$:
\beq
	\bQ_i\sim B_iu^{\hat M_i-1}, \
	\bQ^i\sim B^iu^{-\hat M_i}
\label{asymptQ}
\eeq
where
\begin{align}\label{Mhi}
	\hat M_i=&
	\left\{
	\frac{\Delta -S_1-S_2+2}{2} ,
	\frac{\Delta +S_1+S_2}{2}
		,
	\frac{-\Delta-S_1+S_2+2}{2} ,
	\frac{-\Delta+S_1-S_2}{2} \right\}\ .
\end{align}
From consistency of the above equations we can extract the relation between $\Delta$ and the leading coefficients $A_a,A^a$ in $\bP_a,\bP^a$ appearing in \eq{asymptP}. To do this we expand $Q_{a|i}$, $\bP_a$ etc as (asymptotic) series at large $u$, e.g.
\beq
	Q_{a|i}=u^{N_{ai}}\sum\limits_{n=0}^\infty\frac{B_{ai,n}}{u^n}\ .
\eeq
Then from \eq{QPQ} we find at leading order
\beq
	Q_{a|j}=-iA_aB_j\frac{u^{-\tilde M_a+\hat M_j}}{-\tilde M_a+\hat M_j}\(1+\cO(1/u)\)\ .
\eeq
Plugging this into \eq{QPQ} we see that the coefficients $B_j$ cancel and we get
\beq
	-1=i\sum\limits_{a=1}^4\frac{A_aA^a}{\tilde M_a-\hat M_j},\ \ \ j=1,2,3,4\ .
\eeq
These equations fix the values of $A^{a_0}A_{a_0}$ for $a_0=1,2,3,4$, leading to the important relations
\beq
\label{AAeq}
	A^{a_0}A_{a_0}=i\frac{\prod_j(\tilde M_{a_0}-\hat M_j)}{\prod_{b\neq a_0}(\tilde M_{a_0}-\hat M_b)}\;\;,\;\;a_0=1,2,3,4\;.
\eeq

We see that while the $SO(6)$ charges are encoded in powers of the asymptotics of $\bP$'s, the leading coefficients $A_a$ and $A^a$ contain the remaining charges $S_1,S_2,\Delta$. Equations \eq{asymptP}, \eq{AAeq} supplement the $\bP\mu$ system relations \eq{tPd}-\eq{Papa} and fix the conformal dimension $\Delta$ as a function of $\lambda$ and of the other charges.

Let us finally mention that one can close the equations at the level of $\bQ_i$ (together with $\bQ^i$ which also have one long cut), introducing new functions $\omega_{ij}(u)$ which are analogs of $\mu_{ab}$. The matrix $\omega_{ij}$ is antisymmetric and has unit Pfaffian, but while $\mu_{ab}$ are periodic with long cuts, $\omega_{ij}$ are periodic with short cuts instead.
A related useful property is that while $\mu_{ab}$ has powerlike asymptotics at large real $u$, $\omega_{ij}$ tends to a constant matrix.
The equations relating $\omega$'s with $\bQ$'s read
\beq
\label{Qomsys}
\tilde \bQ_i=\omega_{ij}\bQ^j\;\;,\;\;\tilde \bQ^i=\omega^{ij}\bQ_j
\;\;,\;\;
\tilde\omega_{ij}-\omega_{ij}=\tilde \bQ_i\bQ_j-\bQ_i\tilde\bQ_j\;.
\eeq
where $\omega^{ij}$ is the inverse matrix to $\omega_{ij}$. These are analogs of the $\bP\mu$ system relations we described above.  When these equations are supplemented with constraints on the asymptotics of $\bQ$'s similar to \eq{AAeq}, they form a closed system of equations alternative to the $\bP\mu$-system.

One can also relate $\mu$'s and $\omega$'s with the help of Q-functions with four indices $Q_{ab|ij}$ defined as a determinant,
\beq\label{q4def}
	Q_{ab|ij}=\begin{vmatrix} Q_{a|i}&Q_{a|j} \\ Q_{b|i}&Q_{b|j} \end{vmatrix}\ .
\eeq
Then we have
\beq
\label{moq}
	\mu_{ab}=\frac{1}{2}Q_{ab|ij}^-\omega^{ij}\ .
\eeq

In the next section we will discuss how to change the large $u$ asymptotics to describe the generalized cusp.

\subsection{Adapting the QSC for the cusp anomalous dimension}
\label{sec:QSCcusp}

In this section we will discuss the modifications in the QSC which are needed to describe the generalized quark-antiquark potential. Below we will only discuss the vacuum state, i.e. the Wilson line with $L$ scalar insertions at the cusp (the extension for more general insertions should be straightforward).

The Quantum Spectral Curve equations of \cite{PmuPRL,PmuLong} in $\cN=4$ SYM can be deduced from the TBA equations or the corresponding T- and Y-systems with special analyticity assumptions. In our case the TBA equations for the generalised cusp are almost the same as the original TBA system. The Y-system and T-system equations are exactly the same as for the original problem. Thus it is natural to expect that the QSC equations should also be the same to a large extent. In the TBA there are only two important differences: the extra boundary dressing phase supplementing the BES phase, and the twists which appear as chemical potentials and introduce the angles $\phi,\theta$ into the problem\footnote{There is also an extra symmetry requirement on the Y-functions of the TBA, namely they should be invariant under the exchange of the two wings of the Y-system with a simultaneous reflection $u\to-u$, i.e. $Y_{a,s}(u)=Y_{a,-s}(-u)$, see \cite{Correa:2012hh,Drukker:2012de} for details.}. We do not derive the QSC from the Thermodynamic Bethe ansatz, rather we will put forward and motivate a proposal which is consistent with several highly nontrivial checks, leaving little doubt as to its correctness.

First, we expect to have the same set of Q-functions and auxiliary functions such as $\mu_{ab}$ as in the original problem. All of them will satisfy the same functional relations, for instance the $\bP\mu$-system equations \eq{tPd}-\eq{Papa} or the QQ relations are unchanged. However some analyticity properties will change, as we will discuss below, and in particular the $\bP$-functions acquire an extra cut going from $u=0$ to infinity. In addition, the large $u$ asymptotics clearly need to be modified. Indeed, the twists in the boundary conditions typically correspond to imposing exponential rather than powerlike asymptotics for the Q-functions (see e.g. \cite{PmuLong} and references therein). In our case the angle $\theta$ is naturally related to the $S^5$ part of the geometry, which qualitatively corresponds to the $\bP$-functions, so roughly speaking we expect $\bP_a\sim e^{\pm\theta u}$ at large $u$. Similarly, the angle $\phi$ is associated to $AdS_5$ leading to $\bQ_i\sim e^{\pm\phi u}$. This argument is also supported by the expectation that $\bP$'s and $\bQ$'s should be related in the classical limit to the quasimomenta for $S^5$ and $AdS_5$, correspondingly. Similarly, we expect that $L$ should enter the power in the asymptotics of $\bP$'s, while the power in the asymptotics of $\bQ$'s should contain $\Delta$.

In the original QSC proposal \cite{PmuLong} some guidance to fix the powers in the asymptotics \eq{asymptP}, \eq{asymptQ} came from comparison with the Asymptotic Bethe ansatz (ABA) which can be reproduced from the QSC, and also with the classical spectral curve. For our problem the ABA is also available \cite{Correa:2012hh,Drukker:2012de}, and another piece of information is the all-loop solution of the $\bP\mu$ system to leading order in the near-BPS expansion, based on analytic solutions of the TBA \cite{Gromov:2012eu,PmuPRL,Gromov:2013qga}. In particular these solutions suggest that the large $u$ asymptotics should contain half-integer powers coming from a $\sqrt{u}$ prefactor which the $\bP$'s contain.
However it turns out that there is an important subtlety -- in the near-BPS limit the leading large $u$ coefficient in $\bP_3,\bP_4$ vanishes, making it not straightforward to guess the correct asymptotics even knowing the all-loop result.

The available data indicates that, first, the boundary dressing phase leads to \textit{exponential} rather than powerlike asymptotics in $\mu_{ab}$. This was already observed in \cite{PmuPRL,Gromov:2013qga}. More precisely, we should have
\beq
\label{omexp}
	\omega^{12}\sim {\rm{\const}} \cdot e^{2\pi |u|}, \
	\omega^{13}\sim {\rm{\const}},\
	\omega^{24}\sim {\rm{\const}},\  \ \ \
		u\to\infty
\eeq
while other components of $\omega^{ij}$ become zero at infinity. This translates via \eq{moq} into $e^{\pm 2\pi u}$ asymptotics in some components of the $\mu_{ab}$ matrix.

It remains to fix the powers in the asymptotics of $\bP$'s and $\bQ$'s, and relate their large $u$ expansion coefficients to the charges of the state. To do this
we demanded consistency of the equations \eq{QQP}, \eq{QPQ} expanded at large $u$. This precisely follows the logic described
in the previous section. However, our case turned out to be much more tricky, in particular since some of the twists are the same (e.g. two of the $\bP_a$ functions scale with the same exponent $\sim e^{\theta u}$) there are many subtle cancellations at the first several orders.
It was also convenient at intermediate steps to use \eq{detQ1} as well as the 4th order Baxter-type difference equation on $\bQ_i$ with coefficients built from $\bP_a,\bP^a$ -- this equation follows from \eq{QQP}, \eq{QPQ} (see \cite{Alfimov:2014bwa} for details on its derivation). Finally, already the near-BPS solution suggests that not all four $\bP_a$ are independent, e.g. $\bP_1(u)$ is equal up to a constant to $\bP_2(-u)$.

As a result, we found the following large $u$ asymptotics:
\beqa
\label{cuspas}
\bP_1(u)&=&C\epsilon^{1/2}\;u^{-1/2-L}\; e^{+\theta u} f(+u)\;\;,\;\;f(u)=1+a_1/u+a_2/u^2+a_3/u^3+\dots\\
\nn
\bP_2(u)&=&C\epsilon^{1/2}\;u^{-1/2-L}\; e^{-\theta u} f(-u)\ ,\\
\nn
\bP_3(u)&=&\frac{1}{C}\epsilon^{3/2}\;u^{+3/2+L}\; e^{+\theta u} g(+u)\;\;,\;\;g(u)=1+b_1/u+b_2/u^2+b_3/u^3+\dots\\
\nn
\bP_4(u)&=&-\frac{1}{C}\epsilon^{3/2}\;u^{+3/2+L}\; e^{-\theta u} g(-u)\ .
\eeqa
Here $L$ is the number of scalar insertions at the cusp, while the constant $C$ is unfixed and can be set to $1$ by the rescaling symmetry as discussed below \eq{Presc}, \eq{muresc}.
The coefficients should satisfy
\beq
\label{epsab}
\epsilon^2=\frac{i (\cos \theta -\cos \phi)^2}{2 (L+1) \sin^2 \theta}\;\;,\;\;
a_1-b_1=-\frac{(L+1) (2 \cos \theta  \cos \phi+\cos 2 \theta -3)}{2 \sin\theta (\cos \theta -\cos \phi)}\;.
\eeq
The relation which includes $\Delta\equiv \Gc$ is more involved and we give its full form in Eq. \eq{DLfull}, Appendix \ref{sec:asympDL}. For $L=0$ it reduces to
\beqa
\label{deltaas}
	\Delta^2&=&
	\frac{(\cos\theta-\cos\phi)^3}{\sin\theta\sin^2\phi}\[
	-a_1a_2+a_1b_2-\frac{a_1}{\sin^2\theta}
	+a_1^2\frac{ (1-\cos \theta
   \cos \phi )}{\sin\theta(\cos\theta -\cos \phi )}\right.
	\\ \nn
	&-&\left.\frac{\mathstrut}{\mathstrut} a_2 \cot \theta +a_3-b_3\]\;.
\eeqa
We see that in contrast to the undeformed case we need to expand $\bP$'s up to \textit{fourth} order at large $u$ to extract the conformal dimension! With these asymptotic constraints the $\bP\mu$-system becomes a closed a set of equations fixing the cusp anomalous dimension.

Notice that the asymptotics of $\bP_a$ contains half-integer powers of $u$. Thus $\bP_a$ are not as regular as in the case of local operators and should necessarily have extra cuts. Thus we require the regularity on the plane with only Zhukovsky cuts not for $\bP_a$ (or $\bQ_i$) but for
\beq
\label{defpq}
	\bp_a\equiv\bP_a/\sqrt{u},\ \ \bq_i\equiv\bQ_i/\sqrt{u}\ .
\eeq	
This is an important additional analyticity condition.

Alternatively to the $\bP\mu$-system one can use the $\bQ\omega$ system described in \eq{Qomsys} which is also a closed set of equations provided the proper constraints at large $u$ are imposed. In our case the leading asymptotics of $\bQ_i$ are
\beq
\label{cuspasQ}
	\bQ_1\sim u^{1/2+\Delta}e^{u\phi},\ \bQ_2\sim u^{1/2+\Delta}e^{-u\phi},\
	\bQ_3\sim u^{1/2-\Delta}e^{u\phi},\ \bQ_4\sim u^{1/2-\Delta}e^{-u\phi}\ .
\eeq
The coefficients in their large $u$ expansion are constrained similarly to \eq{cuspas}, \eq{deltaas}, and in particular one can extract from them the R-charge $L$. We give the corresponding relations in Appendix \ref{sec:asympQ}.

Finally, like in the $sl(2)$ sector of the original QSC we have
\beq
\label{Pud}
\bP^1 = - \bP_4\;\;,\;\;
\bP^2 = + \bP_3\;\;,\;\;
\bP^3 = - \bP_2\;\;,\;\;
\bP^4 = + \bP_1,\ \ \mu_{14}=\mu_{23}
\eeq
due to which $\bP^a\bP_a=0$ is satisfied automatically.

It is useful to note that there is a rescaling symmetry under which
\beq
\label{Presc}
	\bP_1\to\alpha\bP_1,\ \bP_2\to\alpha\bP_2,\ \bP_3\to\alpha^{-1}\bP_3,\ \bP_4\to\alpha^{-1}\bP_4,\
\eeq
\beq
\label{muresc}
	\mu_{12}\to\alpha^2\mu_{12},\ \ \ \ \mu_{34}\to\alpha^{-2} \mu_{34}
\eeq
while other $\mu_{ab}$ are not changed ($\alpha$ is a constant). In particular with this rescaling one can set to 1 the constant $C$ appearing in \eq{cuspas}. We also have the $\gamma$-symmetry transformation \cite{Gromov:2014bva,PmuLong} which reads
\beq
\label{gammaP}
	\bP_3\to\bP_3+\gamma\bP_1,\ \bP_4\to\bP_4-\gamma\bP_2,\
\eeq
\beq
\label{gammaMu}
	\mu_{14}\to\mu_{14}-\gamma\mu_{12},\
	\mu_{34}\to\mu_{34}+2\gamma\mu_{14}-\gamma^2\mu_{12}
\eeq
with constant $\gamma$. With this transformation the coefficients in the asymptotics of $\bP$'s will also change, e.g. for $L=0$
\beq
\label{gammabs}
	b_2\to b_2+\frac{C^2\gamma}{\epsilon},\ b_3\to b_3+\frac{C^2\gamma}{\epsilon}a_1,\ \dots
\eeq
The formula \eq{deltaas} for $\Delta$ is invariant under this transformation, as it should be.

As discussed above, from \eq{cuspas} we see that when $\phi\to\theta$ the leading coefficient in $\bP_3,\bP_4$ is proportional to $(\phi-\theta)^{3/2}$ and thus is not visible at the leading order in the near-BPS expansion. The next coefficients $b_1,b_2,\dots$ will scale as $1/(\phi-\theta)$ and thus all $\bP_a$ are of order $\sqrt{\phi-\theta}$,  as expected from the solution found in \cite{Gromov:2013qga}. We will reconstruct this solution in the next section.

The asymptotics discussed in this section constitute our main result. They provide the crucial boundary conditions, thus concluding the reduction of the infinite TBA system of \cite{Correa:2012hh,Drukker:2012de} to the finite set of QSC equations.

In the next sections we will demonstrate the usage of the QSC in several cases. We will compute at all loops the next-to-leading term in the near-BPS expansion, solve the equations numerically and also construct the leading weak coupling solution. All these calculations provide stringent tests of our proposal as well as giving new results.

\section{Near-BPS solution}
\label{sec:nearbps}

In this section we will describe the solution of the QSC in the near-BPS limit $\phi\to\theta$. We will first recover the leading order solution at arbitrary $\theta$ found in \cite{Gromov:2013qga}, and then extend it to the next order. This calculation is quite similar to the iterative solution of the QSC at small spin studied in \cite{Gromov:2014bva}. The main outcome is a prediction for the value of $\Gc$ at order $(\phi-\theta)^2$ to all loops.

\subsection{Leading order}

In the limit $\phi\to\theta$ the generalized cusp anomalous dimension can be written as
\beq
	\Delta=\frac{\cos\phi-\cos\theta}{\sin\phi}\Delta^{(1)}(\phi)
	+\(\frac{\cos\phi-\cos\theta}{\sin\phi}\)^2\Delta^{(2)}(\phi)+\cO((\phi-\theta)^3)\ .
\eeq
The first coefficient, also known as the Bremsstrahlung function, was computed at any coupling in \cite{Correa:2012at,Fiol:2012sg} and later reproduced from integrability in \cite{Gromov:2012eu,Gromov:2013qga} by a direct analytic solution of the TBA in this limit. It reads
\beq
	\Delta^{(1)}(\phi)=\frac{2\phi g}{\sqrt{\pi^2-\phi^2}}
	\frac{I_2\(4\pi g\sqrt{1-\frac{\phi^2}{\pi^2}}\)}{I_1\(4\pi g\sqrt{1-\frac{\phi^2}{\pi^2}}\)}\ .
\eeq
In \cite{Gromov:2013qga} the leading near-BPS solution was obtained from the TBA and linked to the $\bP\mu$-system.
Let us rederive this solution  using solely the information coming from our asymptotics.

The key simplification is that $\bP_a,\tilde\bP_a\sim\sqrt{\phi-\theta}$ are small. This can be seen from our general asymptotics \eq{cuspas},
\eq{cuspasQ}, \eq{Qfullas} where we have to send $\epsilon\sim \phi-\theta\to 0$ meaning that in the near-BPS limit we get
\beq
\label{Pasbps}
\bP_1\sim u^{-1/2-L}	e^{+\theta u}
\;\;,\;\;
\bP_2\sim u^{-1/2-L}	e^{-\theta u}
\;\;,\;\;
\bP_3\sim u^{1/2+L}	e^{+\theta u}
\;\;,\;\;
\bP_4\sim u^{1/2+L}	e^{-\theta u}\ ,
\eeq
and
\beq
\label{Qasbps}
	\bQ^1\sim u^{-1/2-L}	e^{-\theta u}
\;\;,\;\;
\bQ^2\sim u^{-1/2-L}	e^{+\theta u}
\;\;,\;\;
\bQ^3\sim u^{1/2+L}	e^{-\theta u}
\;\;,\;\;
\bQ^4\sim u^{1/2+L}	e^{+\theta u}\ ,
\eeq
Notice that the leading coefficient in $\bP_3$ and $\bP_4$ tends to zero faster than the subleading ones since $a_1-b_1\sim 1/\epsilon$, which 
modifies the expected behaviour at infinity in this limit (and similarly for $\bQ_3,\bQ_4$). Thus we can write the expansion of $\bP$ and $\mu$ as
\beq\label{Pmuser}
	\bP_a=\bP_a^{(0)}+\bP_a^{(1)}+\cO((\phi-\theta)^{5/2}),\ \ \ \
	\mu_{ab}=\mu_{ab}^{(0)}+\mu_{ab}^{(1)}+\cO((\phi-\theta)^{2})
\eeq
where the scaling is
\beq
	\bP_a^{(0)}\sim(\phi-\theta)^{1/2},\ \bP_a^{(1)}\sim (\phi-\theta)^{3/2},
	\ \ \ \ \ \ \  \mu_{ab}^{(0)}\sim 1,\ \ \mu_{ab}^{(1)}\sim(\phi-\theta)\;.
\eeq

From \eq{tmud} we see that at leading order the discontinuity of $\mu_{ab}$ vanishes so $\mu_{ab}^{(0)}$ are periodic entire functions. To fix them we should look in more detail at the functions $Q_{a|i}$ and $Q_{ab|ij}$, using \eq{moq} and our prescription \eq{omexp} which states in particular that $\omega^{12}\sim e^{2\pi |u|},\ u\to\infty$. For $\phi\simeq\theta$ the r.h.s. of \eq{QPQ} is small so $Q_{a|i}$ are periodic functions. At the same time their large $u$ asymptotics should be consistent with that of $\bQ_i$ and $\bP_a$ from \eq{Pasbps}, \eq{Qasbps}, meaning that $Q_{a|i}\simeq u^{N_{ai}}e^{\psi_{ai} u}$
where $\psi_{ai}$ can be equal to $\pm 2\theta$ or to $0$ in our limit. From that we conlude that $Q_{a|i}$ must be constants. Moreover the relation \eq{QPQ},
\beq
	\bP_a=-\bQ^iQ_{a|i}^+\ ,
\eeq
together with \eq{Pasbps}, \eq{Qasbps} means that the only nonzero constants are
\beq
\label{QaiLO}
	Q_{a|i}= \begin{pmatrix}
	0& K_1&0&0\\
	K_2& 0&0&0\\
	0& 0&0&K_3\\
	0& 0&K_4&0\\
	\end{pmatrix}\ .
\eeq
In other words $\bP_a$ and $\bQ_i$ are the same in this limit after a relabeling of their indices (up to a constant factor). This is indeed an expected feature  for a BPS configuration where cancellation between $S^5$ and $AdS_5$ modes
is taking place. Similarly, $\omega^{ij}$ and $\mu^{ab}$ should coincide after the same relabeling of indices.
	
Together with our requirement \eq{omexp} this means that $\mu_{12}=B_0+B_1e^{2\pi u}+B_2e^{-2\pi u}$, $\mu_{13}$ and $\mu_{24}$ are constants, while other $\mu_{ab}$ are zero. Note that since we should have a $u\to-u$ symmetry of the system, of course $\mu_{12}$ should be either even or odd which further constrains these constants. More formally,
from the asymptotics of the $\bP$'s in \eq{cuspas} we see that these functions have the following symmetry under $u\to -u$:
\beq
\label{PSP}
	\bP^a(u)=S^a_{\;\;b}\bP^b(-u),\ \
\eeq
where the matrix $S$ reads\footnote{here one should be careful due to the extra cut in $\bP_a$ going to infinity, and we understand this equation to hold for $\Re(u)>0$}
\beq
	S=\begin{pmatrix}
	0&i&0&0\\
	i&0&0&0\\
	0&0&0&-i\\
	0&0&-i&0
	\end{pmatrix}\ .
\eeq
Due to this we have from the first equation in \eq{tup} together with \eq{PcP}
\beq
\label{muS}
	\mu(-u)=-S^{-1}\mu(u) \chi S \chi
\eeq
(notice also that $S^{-1}=-S$ and $\chi S\chi=-S$).
 Imposing now the symmetry \eq{muS} we get $\mu_{12}=A\sinh(2\pi u)$ and $\mu_{13}=-\mu_{24}$. From ${\rm Pf}(\mu)=1$ we also find that $\mu_{13}=\pm 1$. However with $\mu_{13}=-1$ we found that the equations on the $\bP$'s \eq{tPd} at leading order have no solution consistent with the asymptotics \eq{cuspas}. Thus in summary we get
\beq
\label{muLO}
	\mu_{12}^{(0)}=A\sinh(2\pi u),\ \mu_{13}^{(0)}= 1,\ \mu_{14}^{(0)}=0,\ \mu_{24}^{(0)}=-1,\ \mu_{34}^{(0)}=0
\eeq
where $A$ is a constant. This also implies that at leading order
\beq
\label{om12BPS}
	\omega^{12}={\rm{const}}\cdot\sinh(2\pi u)\ .
\eeq
Therefore the equations on the $\bP$'s \eq{tPd} to leading order take the form
\beqa
\label{LOPmu}
    \tilde\bP_1^{(0)}&=&A\sinh(2\pi u)\bP_3^{(0)}-\bP_2^{(0)}\\ \nn
    \tilde\bP_2^{(0)}&=&A\sinh(2\pi u)\bP_4^{(0)}-\bP_1^{(0)}\\ \nn
    \tilde\bP_3^{(0)}&=&\bP_4^{(0)}\\ \nn
    \tilde\bP_4^{(0)}&=&\bP_3^{(0)}\;.
\eeqa
To solve them let us first introduce some notation. We have a very useful expansion
\beq
	\sinh(2\pi u)e^{+2g\theta(x-1/x)}=\sum\limits_{n=-\infty}^\infty I^{+\theta}_n x^n\;\;,\;\;
\eeq
where
\beq
	I_n^\theta=\frac{1}{2}I_{n}\(4\pi g\sqrt{1-\frac{\theta^2}{\pi^2}}\)\[
	\(\sqrt{\frac{\pi+\theta}{\pi-\theta}}\)^{n}-
	(-1)^n\(\sqrt{\frac{\pi-\theta}{\pi+\theta}}\)^{n}
	\]\;,
\eeq
with $I_n$ being the modified Bessel function. By $x(u)$ we denote the usual Zhukovsky variable which resolves the cut $[-2g,2g]$,
\beq
	x+\frac{1}{x}=\frac{u}{g},\ \ \ |x|\geq 1\ .
\eeq
We also have
\beq
	I_{-n}^\theta=I^{-\theta}_n=(-1)^{n+1} I_n^\theta
\eeq
and let us introduce
\beq
\label{Sdef}
S_+(x)\equiv \sum_{n=1}^\infty I^{+\theta}_n x^{-n}\;\;,\;\;
S_-(x)\equiv \sum_{n=1}^\infty I^{-\theta}_n x^{-n}\;\;.
\eeq
In this notation we have e.g.
\beq
\label{sinsplit}
	S_++\tilde S_-=\sinh(2\pi u)e^{-2 g\theta(x-1/x)}
\eeq
(notice that applying the tilde amounts to flipping $x\to 1/x$). We see that $S_+$ is the part of the Laurent expansion of $\sinh(2\pi u)e^{-2 g\theta(x-1/x)}$ containing negative powers of $x$.
We can alternatively write it as a contour integral
\beq
	S_+(x)=\frac{1}{2\pi i}
	\oint\frac{dy}{x-y}\sinh(2\pi g(y+1/y))e^{-2 g\theta(y-1/y)}
\eeq
where the contour goes counterclockwise around the unit circle.

Focussing on the case $L=0$ we can now write the explicit solution of \eq{LOPmu}:\footnote{This solution is slightly different from the one described in \cite{Gromov:2013qga}, as e.g. the relations \eq{Pud} between $\bP^a$ and $\bP_a$ that we use differ by a sign compared to those used in that paper. The solution given in \cite{Gromov:2013qga} is of course also valid, in the conventions used in that work.}
\beqa
\label{PLO1}
\bP^{(0)}_1&=&B\sqrt{A}\sqrt{u}\;e^{+g\theta(x-1/x)}\sum_{n=1}^\infty I^{+\theta}_n x^{-n}\\
\bP^{(0)}_2&=&B\sqrt{A}\sqrt{u}\;e^{-g\theta(x-1/x)}\sum_{n=1}^\infty I^{-\theta}_n x^{-n}\\
\bP^{(0)}_3&=&\frac{B}{\sqrt{A}}\sqrt{u}\;e^{+g\theta(x-1/x)}\\
\label{PLO4}
\bP^{(0)}_4&=&\frac{B}{\sqrt{A}}\sqrt{u}\;e^{-g\theta(x-1/x)}
\eeqa
where $B\sim\sqrt{\phi-\theta}$ is a constant fixed from asymptotics \eq{cuspas} as
\beq
	B=\sqrt{\frac{-i(\phi-\theta)}{gI_1^\theta}}\ .
\eeq
The constant $A$ is arbitrary and is related to the constant $C$ appearing in the asymptotics \eq{cuspas}, so using the rescaling \eq{Presc}, \eq{muresc} one can set either $A$ or $C$ to 1. One can check that this solution is fully consistent with the asymptotics \eq{cuspas}, noting that, as discussed above, in \eq{cuspas} the leading coefficient in $\bP_3,\bP_4$ vanishes and all $b_i\sim1/(\phi-\theta)$. This solution also reproduces via \eq{deltaas} the known result for $\Delta$ at the leading order in $(\phi-\theta)$,
\beq
\label{DeltaLO}
	\Delta=-2(\phi-\theta)\frac{\phi g}{\sqrt{\pi^2-\phi^2}}
	\frac{I_2\(4\pi g\sqrt{1-\frac{\phi^2}{\pi^2}}\)}{I_1\(4\pi g\sqrt{1-\frac{\phi^2}{\pi^2}}\)}+\cO((\phi-\theta)^2)
	\ .
\eeq
We also translated to our conventions the solution for any $L$ constructed in \cite{Gromov:2013qga} and we present it in appendix \ref{sec:appbps}. Remarkably, the result for $\Gc$ extracted from this solution via our asymptotic relations \eq{cuspas}, \eq{deltaas} perfectly matches the known predictions from TBA found in \cite{Gromov:2013qga} (we have checked this explicitly for the first several values of $L$). This is already a nontrivial check of the proposed large $u$ asymptotics.

\subsection{Next-to-leading order}

Let us now discuss how to solve the $\bP\mu$ system at the next order in $(\phi-\theta)$. The calculation is quite similar to the one done in \cite{Gromov:2014bva} for the small spin expansion, so we will be brief in some cases.

\subsubsection{Constructing $\mu$'s}
\label{sec:mu1}

We will first solve the equation for the correction to $\mu$, which reads
\beq
\label{mu1P0}
	\mu_{ab}^{(1)}(u+i)-\mu^{(1)}_{ab}(u)=\tilde\bP^{(0)}_a\bP^{(0)}_b-\bP^{(0)}_a\tilde\bP^{(0)}_b
\eeq
where $\bP^{(0)}$ are given by \eq{PLO1}-\eq{PLO4} (we understand $\mu$ as functions with short cuts). The function we will actually need is not $\mu$ itself, but rather
\beq
\mu^{\rm reg}_{ab}\equiv \mu_{ab}+\frac{1}{2}\Delta\mu_{ab}
\eeq
where
\beq
	\Delta\mu_{ab}=\tilde\bP_a\bP_b-\bP_a\tilde\bP_b
\eeq
is the discontinuity of $\mu_{ab}$ on the cut on the real axis. Due to the relation $\bP^a\bP_a=0$ one can just as well use $\mu^{\rm reg}$ instead of $\mu$ in the r.h.s. of the $\bP\mu$ system equations \eq{tPd}. The key point is that $\mu^{\rm reg}$
does not have a cut on the real axis and is thus a much nicer function. It should satisfy
\beq
\label{muregeq}
\(\mu^{\rm reg}\)(u+i/2)-
\(\mu^{\rm reg}\)(u-i/2)=\frac{1}{2}\[\Delta\mu(u+i/2)+\Delta\mu(u-i/2)\]\;.
\eeq
Our goal is to solve this equation when the r.h.s. is composed from the leading order $\bP$'s as in \eq{mu1P0}, in particular this means that $\Delta\mu$ only has one cut at $[-2g,2g]$.
The formal solution to this equation can then be written as an integral operator acting on $\Delta\mu$,
\beq
\label{Gammadef}
	\mu^{\rm reg} = \Gamma\cdot\Delta\mu=\oint_{-2g}^{2g}\frac{dv}{4\pi i}
\Gamma_0(u-v)\Delta\mu(v)
\eeq
where the contour goes clockwise around the cut $[-2g,2g]$ and the kernel is
\beqa
\label{Gamma0}
	\Gamma_0(u)&=&\d_u\log\frac{\Gamma(i u+1)}{\Gamma(-i u+1)}
\eeqa
This expression would work well if $\Delta\mu$ decayed powerlike at infinity. A novel feature compared to \cite{Gromov:2014bva} is that we also can have functions with exponential asymptotics $e^{\pm2\theta u}$ in the r.h.s. of \eq{muregeq}. For them the solution is written in a similar way but with a $\theta$-dependent kernel. Namely, if $\Delta\mu(u)=e^{\pm2\theta u}(c/x+\cO(1/x^2))$ we have
\beq
\mu^{\rm reg}=\Gamma_{\pm}\cdot \Delta\mu =
\oint_{-2g}^{2g}\frac{dv}{4\pi i}
\Gamma_{\pm\theta}(u-v)\Delta\mu(v)
\eeq
with the kernels
\beqa
\label{Gammam}
\Gamma_{-\theta}(u)&=&
e^{-2\theta u}
\[
-i e^{-2i\theta}\Phi(e^{-2 i\theta},1,1-iu)
-i e^{+2i\theta}\Phi(e^{+2 i\theta},1,1+iu)
\]\\
\label{Gammap}
\Gamma_{+\theta}(u)&=&
e^{+2\theta u}
\[
-i e^{+2i\theta}\Phi(e^{+2 i\theta},1,1-iu)
-i e^{-2i\theta}\Phi(e^{-2 i\theta},1,1+iu)
\]\ .
\eeqa
Here $\Phi$ is the Hurwitz-Lerch transcendent function\footnote{In Wolfram Mathematica it is the function
\texttt{HurwitzLerchPhi}.}.
Equivalently,
\beq
	\Gamma_{-\theta}(u)=e^{-2\theta u}\sum\limits_{n=1}^\infty
	\[\frac{e^{-2i n\theta}}{u+in}-\frac{e^{2i n\theta}}{u-in}\]\ ,
\eeq
\beq
	\Gamma_{+\theta}(u)=e^{2\theta u}\sum\limits_{n=1}^\infty
	\[\frac{e^{2i n\theta}}{u+in}-\frac{e^{-2i n\theta}}{u-in}\]\ .
\eeq
One final remark is that if we need to solve \eq{muregeq} where $\Delta\mu$ is not decaying at infinity, we subtract the non-decaying part of $\Delta\mu$ and then solve the equation for that part separately.
For example, if $\Delta\mu=x-1/x$ we can write it as
\beq
	\Delta\mu=\(x(u)-\frac{1}{x(u)}-\frac{u}{g}\)+\frac{u}{g}\ .
\eeq
The part in brackets is decaying at infinity, and a particular solution of \eq{muregeq} for the remaining part is found as
\beq
	f(u+i/2)-f(u-i/2)=\frac{1}{2}\[\frac{u+i/2}{g}+\frac{u-i/2}{g}\]
	\Rightarrow
	f(u)=-\frac{i}{2g}u^2\ .
\eeq
So for $\Delta\mu=x-1/x$ we would get
\beq
	\mu^{\rm reg}=\Gamma\cdot\(x(u)-\frac{1}{x(u)}-\frac{u}{g}\)-\frac{i}{2g}u^2\ .
\eeq

As a resut, we can compute in closed form the functions $\mu^{\rm reg}$ at next-to-leading order in $(\phi-\theta)$. Let us denote as in \eq{Pmuser}
\beq
	\mu^{{\rm reg}}=\mu^{{\rm reg}(0)}+\mu^{{\rm reg}(1)}+\cO((\phi-\theta)^2)
\eeq
with
\beq
	\mu^{{\rm reg}(0)}\sim\cO((\phi-\theta)^0),\ \ \ \mu^{{\rm reg}(1)}\sim\cO(\phi-\theta)\ .
\eeq
Then we find
\beqa
\nn
\mu^{{\rm reg}(1)}_{12}&=&B^2A
\[
\sinh(2\pi u)
\Gamma\(
u S_+-u S_-
\)
-\Sigma\(
u S_+^2 e^{+2g\theta (x-1/x)}
\)
+\Sigma\(
u S_-^2 e^{-2g\theta (x-1/x)}
\)
\]
\\ \label{mun12}
\\
\mu^{{\rm reg}(1)}_{13}&=&B^2
\[
-\sinh(2\pi u)
\Sigma\(
u e^{+2g\theta (x-1/x)}
\)
+
\Gamma\(
u S_-+u S_+
\)-2i gI_1^\theta  u
\]
\\
\mu^{{\rm reg}(1)}_{14}&=&B^2
\[
\sinh(2\pi u)\frac{iu^2}{2}
+\Sigma\(
u S_+ e^{+2g\theta (x-1/x)}
\)
+\Sigma\(
u S_- e^{-2g\theta (x-1/x)}
\)
\]
\\
\label{mun24}
\mu^{{\rm reg}(1)}_{24}&=&B^2
\[
-\sinh(2\pi u)
\Sigma\(
u e^{-2g\theta (x-1/x)}
\)
+
\Gamma\(
u S_-+u S_+
\)-2i g I_1^\theta u
\]
\\
\label{mun34}
\mu^{{\rm reg}(1)}_{34}&=&\frac{B^2}{A}
\[
\Sigma\(
u e^{+2g\theta (x-1/x)}
\)-\Sigma\(
u e^{-2g\theta (x-1/x)}
\)
\]
\eeqa
where we use the notation $\Sigma(h)$ to denote a particular solution $f(u)$ of the equation
\beq
	f(u+i)-f(u)=h(u)\ .
\eeq
Explicitly, we have
\beqa
\Sigma\(
u e^{-2g\theta (x-1/x)}
\)&=&
\Gamma_-\(
u e^{-2g\theta (x-1/x)}
\)
\\ \nn
&+&\frac{1}{4} i e^{-2 \theta  u}
   \left(2 \cot \theta  \left(4 g^2
   \theta +u\right)+\frac{1}{\sin^2\theta}\right) \ \ ,
	\\
\Sigma\(
u e^{+2g\theta (x-1/x)}
\)&=&
\Gamma_+\(
u e^{+2g\theta (x-1/x)}
\)
\\ \nn
&-&\frac{1}{4} i e^{2 \theta  u}
   \left(2 \cot\theta \left(u-4
   g^2 \theta \right)-\frac{1}{\sin^2\theta}\right)
\eeqa
\beqa
\Sigma\(
u S_+ e^{+2g\theta (x-1/x)}
\)
&=&
\Gamma_+\(
u S_+ e^{+2g\theta (x-1/x)}
\)-\frac{I_1^\theta}{2} i g \cot \theta e^{+2
   \theta  u}\\
\Sigma\(
u S_+ e^{-2g\theta (x-1/x)}
\)
&=&
\Gamma_-\(
u S_+ e^{-2g\theta (x-1/x)}
\)+\frac{I_1^\theta}{2} i g \cot \theta e^{-2
   \theta  u}	\ .
\eeqa

It's also important that with these corrections to $\mu_{ab}$ the Pfaffian constraint is still satisfied, i.e. reconstructing $\mu_{ab}$ via
$\mu_{ab}=\mu_{ab}^{(0)}+\mu^{{\rm reg}}_{ab}-\frac{1}{2}\Delta\mu_{ab}$ we will get ${\rm{Pf}}(\mu)=1$ to order $\cO(\phi-\theta)$.

Having a particular solution of the equation \eq{muregeq} for $\mu^{\rm reg}$, one could also add to it zero modes, i.e. $i$-periodic entire functions. Due to \eq{moq} and our prescription $\omega^{12}\sim e^{\pm 2\pi u}$ they can be either constants or exponents $e^{\pm 2\pi u}$. In \cite{Gromov:2014bva} the zero modes were important to ensure correct asymptotics. Let us see however that in our case all zero modes can be removed by symmetries to leave the unique solution given above. First, as the $\mu$'s we construct already satisfy the parity constraint \eq{muS}, the zero modes $\mu_{ab}^{z.m.}$ need to satisfy it independently. This immediately shows that the constant zero mode can only appear in $\mu_{13},\mu_{24}$ with $\mu_{13}^{z.m.}=-\mu_{24}^{z.m.}=c_{13}$. Moreover, the exponents $e^{\pm 2\pi u}$ can originate only from $\omega^{12}$ and can come therefore either from the $\cO(\phi-\theta)$ correction to $\omega^{12}$ or from the correction to $Q_{ab|12}$. In the first case the correction to $\omega^{12}$ will multiply the leading order $Q_{ab|12}$ and thus exponents will appear only in $\mu_{12}$. They are then restricted by \eq{muS} to appear only in the combination proportional to $\sinh(2\pi u)$. In the second case the leading order $\omega^{12}$ will multiply the correction to $Q_{ab|12}$ so again the exponents will appear as $\sinh(2\pi u)$. By arguments similar to those leading to \eq{QaiLO}, we expect the constant part of $Q_{a|i}$ (in the large $u$ expansion) at order $\cO(\phi-\theta)$ to be
\beq
\label{QaiNLO}
	Q_{a|i}^{\rm{const}}= \begin{pmatrix}
	0& K_1+(\phi-\theta) N_1&0&(\phi-\theta)M_1\\
	K_2+(\phi-\theta) N_2& 0&(\phi-\theta) M_2&0\\
	0& (\phi-\theta) M_3&0&K_3+(\phi-\theta) N_3\\
	(\phi-\theta) M_4& 0&K_4+(\phi-\theta) N_4&0\\
	\end{pmatrix}
\eeq
where for complete generality at order $(\phi-\theta)$ we introduced constants in all components of $Q_{a|i}$ for which the twists $\pm\phi,\pm\theta$ in the asymptotics cancel in the near-BPS limit. From \eq{QaiNLO} we find that out of $Q_{ab|12}$ we will have a nonzero correction to the constant part only for $(a,b)=(1,2),(1,4),(2,3)$. Thus a zero mode proportional to $\sinh(2\pi u)$ could appear only in these components of $\mu_{ab}$. In summary, we find the following possible zero modes:
\beq
	\mu_{12}^{z.m.}=c_1 \sinh(2\pi u),\ \mu_{13}^{z.m.}=c_2,\ \mu_{14}^{z.m.}=c_3\sinh(2\pi u),\ 
	\mu_{24}^{z.m.}=-c_2,\ \mu_{34}^{z.m.}=0
\eeq
where $c_i$ are constants of order $\cO(\phi-\theta)$. From the Pfaffian constraint we get $c_2=0$. Moreover as $\mu_{12}^{(0)}\propto\sinh(2\pi u)$ the constant $c_1$ can be set to zero by a rescaling transformation \eq{muresc} with $\alpha=1+{\rm{const}}\cdot (\phi-\theta)$. Finally we can set $c_2$ to zero by making a $\gamma$ transformation \eq{gammaMu} with $\gamma={\rm{const}}\cdot (\phi-\theta)$ as then this zero mode will cancel against the leading order part of $\mu_{12}$ (notice that $\mu_{34}$ will change under this $\gamma$ transformation only at order $(\phi-\theta)^2$ which is irrelevant for us). Thus we find that the solution for $\mu_{ab}$ at NLO given above is completely general.

For $\mu_{12}$ one can make yet another test of our prescription \eq{omexp}, as its exponential part is nonzero already at leading order in the near-BPS expansion. Let us use \eq{moq}
\beq
\label{moq2}
	\mu_{ab}=\frac{1}{2}Q^-_{ab|ij}\omega^{ij}
\eeq
where as at large $u$ $Q_{12|12}\sim u^{2\Delta}$ we can expand
\beq
\label{q12om}
	Q_{12|12}^-\omega^{12}\sim \const\cdot e^{2\pi u} (1+2\Delta\log u+\cO((\phi-\theta)^2)\ .
\eeq
This gives a prediction for the ratio of the coefficients of the $\log u$ term and the leading term. In our results for $\mu_{12}$  the logarithmic part comes only from $\Gamma\cdot(uS_+-uS_-)$ appearing in \eq{mun12}, so the coefficient of $\log u$ is determined by the $1/u$ term in $(uS_+-uS_-)$ and is straightforward to evaluate. We find
\beq
\label{mu12exp}
	\mu_{12}\sim\frac{1}{2}e^{2\pi u} A\(1+2\gamma \log u+\cO((\phi-\theta)^2)\)+\dots
\eeq
where the coefficient $\gamma$
precisely matches the expression \eq{DeltaLO} for $\Delta$ at leading order in $(\phi-\theta)$.
Thus our result \eq{mu12exp} agrees with the prediction \eq{q12om}.

In the next section we will use the correction to $\mu_{ab}$ to construct $\bP_a$ at the next-to-leading order.

\subsubsection{Constructing $\bP$'s}

Having found the correction to $\mu_{ab}$ we can now proceed and compute the correction to $\bP_a$. It is convenient to parameterize them as
\beqa
\label{POE1}
\bP_1^{(1)}&=&{\bf p}^+ \(O_1(u)+E_1(u)\)\\
\bP_2^{(1)}&=&{\bf p}^- \(O_1(u)-E_1(u)\)\\
\bP_3^{(1)}&=&{\bf p}^+ \(E_3(u)+O_3(u)\)\\
\label{POE4}
\bP_4^{(1)}&=&{\bf p}^- \(E_3(u)-O_3(u)\)
\eeqa
where
\beq
{\bf p}^\pm=\sqrt u \;e^{\pm g\theta (x-1/x)}
\eeq
The functions $E_1,E_3,O_1,O_3$ then have powerlike rather than exponential asymptotics at large $u$. From \eq{cuspas} we find that $O_1,O_3$ are odd, while $E_1,E_3$ are even and at large $u$
\beq
\label{EOas}
	E_1\sim 1/u^2,\ \ \ O_1\sim 1/u,\ \ \ E_3\sim u^0,\ \ \ O_3\sim u\ .
\eeq
Plugging this parameterization into the $\bP\mu$ system equations \eq{tPd}, we find the following equations on these functions:
\beqa
\label{OEeq1}
\tilde O_1+O_1&=&\frac{1}{2}
\[
\frac{A\sinh(2\pi u) \bP_3^{(1)}+\delta_1}{{\bf p}^-}
+
\frac{A\sinh(2\pi u) \bP_4^{(1)}+\delta_2}{{\bf p}^+}
\]\\
\tilde E_1-E_1&=&\frac{1}{2}
\[
\frac{A\sinh(2\pi u) \bP_3^{(1)}+\delta_1}{{\bf p}^-}
-
\frac{A\sinh(2\pi u) \bP_4^{(1)}+\delta_2}{{\bf p}^+}
\]\\
\label{E3eq}
\tilde E_3-E_3&=&\frac{1}{2}
\[
\frac{\delta_3}{{\bf p}_-}+
\frac{\delta_4}{{\bf p}_+}
\]\\
\label{OEeq4}
\tilde O_3+O_3&=&\frac{1}{2}
\[
\frac{\delta_3}{{\bf p}_-}-
\frac{\delta_4}{{\bf p}_+}
\]
\eeqa
where we denoted
\beq
	\delta_a=\mu_{ab}^{{\rm reg}(1)}\bP^{b(0)}\ .
\eeq	
Since we have computed $\mu_{ab}^{{\rm reg}(1)}$ in the previous section, $\delta_a$ are some explicitly known functions. As in \cite{Gromov:2014bva} we can write a particular solution of these equations as some integral operator acting on the r.h.s. First, the equation
\beq
	\tilde f+f=h\ ,
\eeq
where necessarily $\tilde h=h$, is solved by
\beq
\label{Hsol}
	f=H\cdot h=-\oint\frac{dv}{4\pi i}\frac{\sqrt{u-2g}}{\sqrt{v-2g}}
	\frac{\sqrt{u+2g}}{\sqrt{v+2g}}\frac{1}{u-v}h(v)\
	 ,
\eeq
where the integral is clockwise around the cut $[-2g,2g]$. The solution found this way has constant asymptotics at infinity. Similarly, for the equation
\beq
	\tilde f-f=h\ ,
\eeq
with $\tilde h=-h$, a solution decaying at infinity can be written as
\beq
\label{Ksol}
	f=K\cdot h=\oint\frac{dv}{4\pi i}\frac{1}{u-v}
	h(v)
	\ .
\eeq
To the particular solutions \eq{Ksol}, \eq{Hsol} we can also add zero modes, i.e. solutions of the same equations with zero right-hand side.
Then we get
\beqa
	O_3&=&\frac{1}{2}H\cdot \[\frac{\delta_3}{{\bf p}_-}-
\frac{\delta_4}{{\bf p}_+}\]+C_1(x-1/x)\ , \\
	E_3&=&\frac{1}{2}K\cdot \[\frac{\delta_3}{{\bf p}_-}+
\frac{\delta_4}{{\bf p}_+}\]+ C_2
\eeqa
where the constants $C_1,C_2$ parameterize the most general zero modes that do not violate the asymptotics \eq{EOas}.
Now we can compute $\bP_3^{(1)},\bP_4^{(1)}$ and then solve the equations on $E_1,O_1$:
\beq
	O_1=\frac{1}{2}K\cdot
\[
\frac{A\sinh(2\pi u) \bP_3^{(1)}+\delta_1}{{\bf p}^-}
-
\frac{A\sinh(2\pi u) \bP_4^{(1)}+\delta_2}{{\bf p}^+}
\]\ ,
\eeq
\beq
	E_1=\frac{1}{2}H\cdot\[
\frac{A\sinh(2\pi u) \bP_3^{(1)}+\delta_1}{{\bf p}^-}
+
\frac{A\sinh(2\pi u) \bP_4^{(1)}+\delta_2}{{\bf p}^+}
\]\ .
\eeq
Having found $E_n$ and $O_n$, we obtain the $\bP$-functions from \eq{POE1}-\eq{POE4}. Then we fix the constants $C_1,C_2$ by imposing the asymptotic constraints \eq{cuspas}, \eq{epsab}.

From the corrected $\bP$'s that we have now computed we can finally extract the correction to the conformal dimension $\Delta$ at all loops. We will present this result in the next section.

\subsubsection{Final result}
From the next-to-leading order solution of the $\bP\mu$ system we constructed above, we obtain a new prediction for the generalized cusp anomalous dimension at order $(\phi-\theta)^2$.
To compare with the literature we found it convenient to bring our result to the form
\beq
\label{dexpphi}
	\Delta=\frac{\cos\phi-\cos\theta}{\sin\phi}\Delta^{(1)}(\phi)
	+\(\frac{\cos\phi-\cos\theta}{\sin\phi}\)^2\Delta^{(2)}(\phi)+\cO((\phi-\theta)^3)
\eeq
so that at each order we have a nontrivial function of $\phi$ .
Our all-loop result reads
\beqa\nn
\label{D2res}
	\Delta^{(2)}(\phi)&=&-\frac{1}{2}\oint\frac{du_x}{2\pi i}\oint\frac{du_y}{2\pi i}
	u_xu_y
	\left[D_+\Gamma_{+\phi}(u_x-u_y)\right.
+
D_0\Gamma_{0}(u_x-u_y)\left.
	+D_-\Gamma_{-\phi}(u_x-u_y)
		\right]
		\\
\eeqa
where both integrals run clockwise around the cut $[-2g,2g]$ and
\beqa\nn
	D_+&=&
	\frac{iS_+(y)e^{-2g\phi x+2g\phi/x+2g\phi y-2g\phi/y}}{g^3I_1^\phi}
	\\ \nn
	&\times&
	\(-\frac{2S_+(y)}{gI^\phi_1}
	\right.
	-
	\frac{2S_+(x)e^{4g\phi x-4g\phi/x}}{gI^\phi_1}
	+\frac{2  y }
	{y^2-1}
	+\left.\frac{2   x}
	{x^2-1}
	+\frac{I^\phi_2 x S_+(y)}
	{(I^\phi_1)^2(x^2-1)}
	\)\ \ ,
	\\
	D_0&=&
	\frac{2iS_+(y)}{g^3I_1^\phi}\(\frac{S_+(x)}{gI^\phi_1}
	-\frac{I^\phi_2 S_+(x)}{(I^\phi_1)^2(x^2-1)}
-
	\frac{2 x^2}{(x+1/x)(x^2-1)}\)\ \ ,
\\ \nn
	D_-&=&\frac{iI^\phi_2}{g^3 (I^\phi_1)^3}
	\frac{x (S_+(x))^2e^{2g\phi x-2g\phi/x-2g\phi y+2g\phi/y}}
	{(x^2-1)}
\eeqa
We recall that $S_+$ was defined in \eq{Sdef}, while the kernels $\Gamma$ entering \eq{D2res} are given in \eq{Gamma0}, \eq{Gammam}, \eq{Gammap}.

It is also interesting to consider a further limit when $\theta$ is set to zero and $\phi$ is small. This corresponds to an expansion near the straight Wilson line. The first two terms in the series \eq{dexpphi} scale as
\beq
	\Delta^{(1)}\sim \frac{2gI_2(4\pi g)}{\pi I_1(4\pi g)}\phi+\cO(\phi^3),\ \ \ \
	\Delta^{(2)}\sim \phi^2 f_2(g)+\cO(\phi^4)
\eeq
where the function $f_2(g)$ is found by expanding our result \eq{D2res} and is given explicitly in Appendix \ref{sec:phi0}. Expanding also the prefactors in \eq{dexpphi} we can write $\Delta$ as a series in powers of $\phi^2$,
\beq
	\Delta=-\phi^2\frac{gI_2(4\pi g)}{\pi I_1(4\pi g)}
	+
	\phi ^4 \left[\frac{1}{4}f_2(g)+\frac{2 g^2}{\pi ^2}-\frac{2 g^2 I_2(4  \pi g){}^2}{\pi
   ^2 I_1(4  \pi g){}^2}-\frac{(24+\pi ^2) g I_2(4  \pi g)}{12 \pi ^3 I_1(4  \pi g
   )}\right]+\cO(\phi^6)
\eeq

Let us now discuss several checks of our main result \eq{D2res} at weak coupling. It is straightforward to expand it for $g\to 0$ simply by expanding the integrand in \eq{D2res} at weak coupling and taking the residue at $u_x,u_y=0$. Then we can make a test against perturbative predictions known up to four loops. In general the structure at weak coupling is expected to be
\beq
	\Delta=\sum\limits_{n=1}^\infty \gamma_n(\theta,\phi)g^{2n}
\eeq
with
\beq
\label{dstruc}
	\gamma_n(\theta,\phi)=\sum_{k=1}^n\(\frac{\cos\phi-\cos\theta}{\sin\phi}\)^k\gamma_{n}^{(k)}(\phi)\;.
\eeq
 Our all-loop result allows to compute all coeficients $\gamma_{n}^{(2)}(\phi)$ in this expansion. Notice that at each loop order only a finite number of terms in the near-BPS expansion contribute, e.g. the two-loop result is completely determined by the first two terms in \eq{dexpphi}. For arbitrary $\phi$ and $\theta$ the anomalous dimension was computed directly up to two loops \cite{Makeenko:2006ds,Drukker:2011za} giving
\beqa
\label{gamma1loop}
	\gamma_1^{(1)}(\phi)&=&2\phi,
	\\
	\gamma_2^{(1)}(\phi)&=&\frac{4}{3}\phi(\phi^2-\pi^2),
	\\ \label{dweak2}
	\gamma_2^{(2)}(\phi)&=&2 i \phi  \[\text{Li}_2(e^{2 i \phi })-\text{Li}_2(e^{-2 i \phi })\]
	-2 \[\text{Li}_3(e^{2 i \phi })+\text{Li}_3(e^{-2 i \phi })\]+4
   \zeta (3)
\eeqa
and in \cite{Bajnok:2013sya} this data was reproduced from the TBA. We found that the weak coupling expansion of our result perfectly matches the prediction \eq{dweak2}.

The cusp anomalous dimension was also computed to four loops in \cite{Correa:2012nk,Henn:2013wfa}, giving a prediction for the coefficients $\gamma_{3}^{(2)}(\phi),\gamma_{4}^{(2)}(\phi)$ which our result should reproduce. Indeed we found a perfect match with the perturbative data.
The predictions of \cite{Henn:2013wfa} are written in terms of harmonic polylogarithms, but match the expansion of our
result\footnote{we checked this numerically for some particular values of $\phi$}
which does not include more complicated functions than ${\text{Li}}_n$. At three loops our result gives
\beqa
\gamma_3^{(2)}(\phi)&&
=
	24 \left[\text{Li}_5(e^{-2 i \phi
   })+\text{Li}_5(e^{2 i \phi })\right]
		-18 i \phi  \left[\text{Li}_4(e^{2 i \phi
   })-\text{Li}_4(e^{-2 i \phi })\right]
	\\ \nn &&
-4 \phi ^2 \[\text{Li}_3(e^{-2 i \phi })+\text{Li}_3(e^{2 i \phi
   })\]
	+\frac{4}{3} i (\pi -\phi ) (\phi +\pi ) \phi  \left[\text{Li}_2(e^{2 i \phi
   })-\text{Li}_2(e^{-2 i \phi })\right]
	\\ \nn &&
	+\frac{8}{3} \left(\phi ^2-\pi ^2\right) \phi ^2
	\[
	\log (1-e^{2 i \phi
   })+\log (1-e^{-2 i \phi
   })
	\]
	+8 \left(\zeta (3) \phi ^2-6 \zeta
   (5)\right)
\eeqa
while at four loops
\beqa
\gamma_4^{(2)}(\phi)&&
=
	-280 \[\text{Li}_7(e^{2 i \phi})+\text{Li}_7(e^{-2 i \phi})\]
	+190 i \phi \[
   \text{Li}_6(e^{2 i \phi })-\text{Li}_6(e^{2 i \phi })
	\]
	\\ \nn &&
	+\left(44 \phi ^2+\frac{16 \pi ^2}{3}\right)
	\[\text{Li}_5(e^{2 i \phi})+\text{Li}_5(e^{-2 i \phi})\]
	\\ \nn &&
	+\frac{4}{3} i \phi  \left(11 \phi ^2-17 \pi ^2\right)
	\[\text{Li}_4(e^{2 i \phi})-\text{Li}_4(e^{-2 i \phi})\]
	\\ \nn &&
	+\frac{8}{9} \left(18 \phi ^4-21 \pi ^2 \phi
   ^2+\pi ^4\right)
	\[\text{Li}_3(e^{2 i \phi })+\text{Li}_3(e^{-2 i \phi })\]
	\\ \nn &&
	-\frac{4}{9} i \left(15 \phi ^5-22 \pi ^2 \phi ^3+7 \pi ^4 \phi \right)
   \[\text{Li}_2(e^{2 i \phi })-\text{Li}_2(e^{-2 i \phi })\]
		\\ \nn &&
	+\frac{40}{9}
   \left(\phi ^3-\pi ^2 \phi \right)^2
	\[
	\log (1-e^{2 i \phi
   })+\log (1-e^{-2 i \phi
   })
	\]
	\\ \nn &&
	+
		16 \zeta (3) \phi ^4-\frac{8}{3} \left(4 \pi ^2 \zeta (3)+33 \zeta (5)\right) \phi ^2-\frac{16}{9} \left(\pi
   ^4 \zeta (3)+6 \pi ^2 \zeta (5)-315 \zeta (7)\right)
\eeqa
In fact it is clear that at any loop order our result would generate ${\rm Li}_n$ at most. The reason is that when evaluating the integral \eq{D2res} by residues the most complicated functions that can appear are the ${\rm Li}_n(e^{\pm 2i\phi})$ coming from expansion of the kernels \eq{Gammam}, \eq{Gammap}.
As a further example we computed the novel five- and six-loop coefficients, given in Eq. \eq{our5loop} (Appendix \ref{sec:weak5}). We attach a Mathematica notebook which allows to reproduce these results and also systematically expand our all-loop result at weak coupling.

Thus at weak coupling our result matches known predictions to four loops, which serves as a deep test of the proposed Quantum Spectral Curve equations and of our near-BPS calculation.

\section{Numerical solution}
\label{sec:num}

The formulation of the problem in terms of the QSC
allows for an efficient numerical analysis of $\Gc$ at finite coupling. A highly precise  and fast converging numerical method for solving the original QSC for local operators was proposed in \cite{Gromov:2015wca}. Here we will describe how to modify it in the present case, and then demonstrate several applications. We will focus on the case $L=0$, but
we expect the discussion in this section should be valid for general $L$ with minor changes.

\subsection{The numerical algorithm}

The first step is to parameterize the $\bP$-functions in terms of the Zhukovsky variable $x(u)$. The only difference with \cite{Gromov:2015wca} is that these functions now have exponential asymptotics, but they still have only one cut. Due to this, after extracting their exponential and leading powerlike asymptotics
like in \eq{cuspas}
they become power series in $1/x$ convergent everywhere on the main sheet. Thus we can approximate the functions $f(u)$ and $g(u)$ appearing in
\eq{cuspas} as
\beq
\label{fgexp}
	f(u)=1+\sum_{n=1}^M \frac{c_{1,n}}{x^n},\ \ \ g(u)=1+\sum_{n=1}^M \frac{c_{2,n}}{x^n}
\eeq
where $M$ is some large cutoff\footnote{
in practice $M\sim 30$ is enough to get at least $10$ digits precision for $0<g<0.85$}. Then we build $\bP$'s from \eq{cuspas} (where we set the constant $C$ to 1) in terms of the coefficients $c_{1,n},\;c_{2,n}$ which are the main parameters in our algorithm.\footnote{We also fix the $\gamma$-symmetry \eq{gammaP}, \eq{gammaMu} by fixing the coefficient $b_2$ appearing in the asymptotics \eq{cuspas} to be zero using \eq{gammabs}.}

The next step is to close the equations which will give constraints fixing the values of these coefficients. For that we construct the functions $Q_{a|i}$ defined by \eq{Qai} in terms of $\bP$-functions,
\beq
\label{Qai2}
{ Q}_{a|i}^+-{ Q}_{a|i}^-=-
	\bP_a
	\bP^b
	{ Q}_{b|i}^+\ \ .
\eeq
As in \cite{Gromov:2015wca} we first solve this equation analytically at large $u$, plugging the asymptotic expansion truncated at some cutoff $K$,
\beq
	Q_{a|i}\simeq e^{(\theta_a +\phi_i )u}u^{N_{ab}}\sum_{n=1}^K B_{ai,n}/u^n\ ,
\eeq
into \eq{Qai2} and obtaining the coefficients $B_{ai,n}$ in terms of $c_{1,n},\; c_{2,n}$ (here $\theta_a=\pm \theta$ and $\phi_i=\pm \phi$). It is important here to account for several cancellations taking place due to the asymptotics \eq{cuspas}. As a result we get a good numerical approximation for $Q_{a|i}(u)$ when $u$ has  sufficiently large imaginary part. Then we use the exact equation \eq{Qai2} to decrease the imaginary part of $u$ and eventually obtain the functions $Q_{a|i}$ in the vicinity of the real axis, when $u\in[-2g+i/2,2g+i/2]$.

Now we can build the $\bQ$-functions on the cut $[-2g,2g]$ via \eq{QPQ},
\beq
	\bQ_i=-\bP^aQ_{a|i}^+\ ,
\eeq
and since $Q_{a|i}^+$ do not have a cut on the real axis we also obtain $\tilde\bQ_i$ on the cut as
\beq
	\tilde\bQ_i=-\tilde\bP^aQ_{a|i}^+\ .
\eeq

The final step is to close the equations in terms of $\bQ_i,\tilde\bQ_i$ and find the free coefficients $c_{1,n}$ and $c_{2,n}$. For that we use the very convenient trick proposed originally in \cite{Gromov:2015vua}. Let us discuss it in some detail as this is a crucial part of the calculation. We start by noticing that $\bQ_i(u)$ and $\bQ_i(-u)$ should satisfy the same 4th order difference equation following from \eq{Qai}, \eq{QPQ} with coefficients built from $\bP$-functions as the equation is symmetric under $u\to -u$. As we discussed in section \ref{sec:QSCcusp}, Eq. \eq{defpq}, it is simpler to work with
\beq
\label{defq}
	\bq_i(u)=\bQ_i(u)/\sqrt{u}\ .
\eeq
Then we have $\bq_i(u)=\Omega^j_i(u)\bq_j(-u)$ where $\Omega^j_i(u)$ are some $i-$periodic functions. As $\bQ_i$ have a definite asymptotics with prescribed exponential part \eq{cuspasQ}, all $\Omega^j_i(u)$ become constant at large $u$ and furthermore only a few of them are nonzero at infinity, namely $\Omega^1_2,\;\Omega^2_1,\;\Omega^4_3,\;\Omega^3_4$.
We also know that $\tilde \bq_i(u)=\omega_{ij}(u)\chi^{jk}\bq_k(u)$ where $\omega_{ij}$ are $i$-periodic. Combining these relations we find
\beq
\label{qgluegen}
	\tilde\bq_m(u)=\alpha^i_m(u)\bq_i(-u),\ \ m=1,2,3,4
\eeq
where $\alpha^i_m=\omega_{mj}\chi^{jk}\Omega^i_k$ are $i$-periodic (being built from periodic functions) and moreover analytic since $\tilde\bq_i(u)$ and $\bq_i(-u)$ are analytic in the lower half-plane. In addition to this, most of the functions $\alpha^i_m$ are equal to zero, because according to our prescription \eq{omexp} from section \ref{sec:QSCcusp} the only nonzero components of $\omega_{ij}$ at infinity are $\omega_{34}\sim {\rm{\const}}\cdot
e^{2\pi |u|}$ and $\omega_{13},\omega_{24}\sim {\rm{\const}}$. Using also that most components of $\Omega^j_i$ are zero at large $u$ we get from \eq{qgluegen} the following equations (it's enough for us to consider only $\bq_1,\bq_4$)
\beqa
\label{qabc}
	\tilde\bq_1(u)&=&s_1\bq_1(-u)\\ \nn
	\tilde\bq_4(u)&=&(ae^{2\pi u}+be^{-2\pi u}+c)\bq_1(-u)+s_4\bq_4(-u)
\eeqa
where $s_1,s_4,a,b,c$ are constants, and moreover $a$ and $b$ are nonzero as $\Omega^1_2$ and the exponential part of $\omega_{34}$ are nonzero at infinity. Applying tilde to the first equation we also get
\beq
	\bq_1(u)=s_1\tilde\bq_1(-u)=(s_1)^2\bq_1(u)
\eeq
so $(s_1)^2=1$. Similarly from the second equation we find $(s_4)^2=1$ as well as
\beqa
	as_1+bs_4&=&0\\ \nn
	bs_1+as_4&=&0\\ \nn
	cs_1+cs_4&=&0\ .
\eeqa
This system has two solutions: either
\beq
\label{solabc1}
	s_1=s_4,\ a=-b,\ c=0
\eeq
or
\beq
	s_1=-s_4,\ a=b,\ \text{ and }c\text{ is arbitrary}.
\eeq
By comparing to the leading near-BPS solution where $\omega^{12}\propto\sinh(2\pi u)$ (see Eq. \eq{om12BPS}), we see that the first option is the correct one. It remains only to fix the sign of $s_1$. For that let us consider the explicit solution \eq{PLO1}-\eq{PLO4} for $\bP_a$ in the near-BPS limit. We see that for $\bp_a=\bP_a/\sqrt{u}$ we have
\beq
	\tilde\bp_3(u)=\bp_3(-u)
\eeq
As in the near-BPS limit we expect to identify $\bq_1$ and $\bp_3$, comparing this relation with the first equation in \eq{qabc} we see that we should choose $s_1=+1$.

In summary, we get a remarkably simple set of equations:
\beqa
\label{qg1}
	\tilde\bq_1(u)&=&\bq_1(-u)\\
\label{qg2}
	\tilde\bq_4(u)&=&A\sinh(2\pi u)\bq_1(-u)+\bq_4(-u)
\eeqa
where $A$ is a constant and we recall that in our notation $\bq_i(u)=\bQ_i(u)/\sqrt{u}$. These are the key equations which are enough to close the system. Let us stress that they are exact and are not restricted to large $u$ or near-BPS limit. In particular, similarly to \cite{Gromov:2015vua} these equations should be useful for a systematic weak coupling solution. With this approach we can completely avoid computing $\omega_{ij}$ as we are able to close the system using various Q-functions only. Notice also that in \cite{Gromov:2015vua} the resulting equations were similar but coefficients in the r.h.s. were all constant, while here we also have $\sinh(2\pi u)$.

Now, finally, as we know $\bQ_i$ and $\tilde\bQ_i$ on the cut, we can evaluate both sides of \eq{qg1}, \eq{qg2} at sampling points $u_k$ on the cut, and minimize the difference between them. More precisely, we can express the constant $A$ from \eq{qg2} as
\beq
	A=\frac{\tilde\bq_4(u)-\bq_4(-u)}{\bq_1(-u)\sinh(2\pi u)}
\eeq
and we build a function which on the true solution of the QSC should be zero\footnote{As in \eq{defF} we have $\sinh(2\pi u_k)$ in denominator we should make sure the sampling points do not include $u_k=0$. We choose $N$ sampling points as $u_k=2gz_k$ where $z_k$ are zeros of the $N$-th Chebyshev polynomial $T_N(z)$.}:
\beq
\label{defF}
	F=\sum_k|\tilde\bq_1(u_k)-\bq_1(-u_k)|^2
	+{\rm{Var}}\[\frac{\tilde\bq_4(u_k)-\bq_4(-u_k)}{\bq_1(-u_k)\sinh(2\pi u_k)}\]
\eeq
where ${\rm{Var}}$ denotes the variance, i.e. measures the deviation of the function from a constant
\footnote{${\rm{Var}}\[f_k\]=\sum_k|f_k-\hat f|^2$ where $\hat f$ is the average of all elements $f_k$.}.
Thus we have reduced the problem to minimization of $F$ which is a function of our main parameters $c_{1,n},\;c_{2,n}$. It's easy to see that $F$ can be written as the norm of a $2N$-dimensional vector where $N$ is the number of sampling points. Therefore to find its minimum we can use the iterative Levenberg-Marquardt algorithm (an improved version of Newton's method) as in \cite{Gromov:2015wca}. It converges rather fast and robustly, giving the values of coefficients $c_{1,n},\;c_{2,n}$. Now we can reconstruct the $\bP$'s and compute the anomalous dimension from e.g. \eq{deltaas}.

\subsection{Results}

\begin{table}{
$$
\begin{array}{||l|l||l|l||l|l||l|l||}
\hline
 g & \Gc(g) &  g & \Gc(g) &  g & \Gc(g) &  g & \Gc(g) \\ \hline
 0.0125 & 0.000138062 & 0.025 & 0.000550881 & 0.0375 & 0.0012344 & 0.05 & 0.00218203 \\
 0.0625 & 0.00338487 & 0.075 & 0.00483202 & 0.0875 & 0.00651094 & 0.1 & 0.00840784 \\
 0.1125 & 0.010508 & 0.125 & 0.0127963 & 0.1375 & 0.0152575 & 0.15 & 0.0178762 \\
 0.1625 & 0.0206379 & 0.175 & 0.0235283 & 0.1875 & 0.0265342 & 0.2 & 0.0296431 \\
 0.2125 & 0.0328434 & 0.225 & 0.0361248 & 0.2375 & 0.0394776 & 0.25 & 0.0428933 \\
 0.2625 & 0.0463641 & 0.275 & 0.0498834 & 0.2875 & 0.053445 & 0.3 & 0.0570437 \\
 0.3125 & 0.0606747 & 0.325 & 0.0643342 & 0.3375 & 0.0680183 & 0.35 & 0.0717242 \\
 0.3625 & 0.0754492 & 0.375 & 0.0791908 & 0.3875 & 0.0829471 & 0.4 & 0.0867164 \\
 0.4125 & 0.0904971 & 0.425 & 0.0942879 & 0.4375 & 0.0980876 & 0.45 & 0.101895 \\
 0.4625 & 0.10571 & 0.475 & 0.109532 & 0.4875 & 0.113359 & 0.5 & 0.117191 \\
 0.5125 & 0.121027 & 0.525 & 0.124868 & 0.5375 & 0.128713 & 0.55 & 0.132561 \\
 0.5625 & 0.136413 & 0.575 & 0.140267 & 0.5875 & 0.144124 & 0.6 & 0.147984 \\
 0.6125 & 0.151845 & 0.625 & 0.155709 & 0.6375 & 0.159575 & 0.65 & 0.163442 \\
 0.6625 & 0.167312 & 0.675 & 0.171182 & 0.6875 & 0.175054 & 0.7 & 0.178928 \\
 0.7125 & 0.182803 & 0.725 & 0.186679 & 0.7375 & 0.190556 & 0.75 & 0.194434 \\
 0.7625 & 0.198313 & 0.775 & 0.202193 & 0.7875 & 0.206074 & 0.8 & 0.209955 \\
 0.8125 & 0.213838 & 0.825 & 0.217721 & 0.8375 & 0.221605 & 0.85 & 0.22549 \\
 \hline
\end{array}
$$
\caption{Numerical data used for the plot in Fig. \protect\ref{fign}. We give the values of $\Gc$ at finite coupling for $\phi=\pi/4,\;\theta=4\pi/10$. 
Precision is decreased to fit the page. The full data set is available as attachment to the
arXiv submission.\label{numtab}}
}
\end{table}

\FIGURE[t]{
\la{fign}
\begin{tabular}{cc}
\includegraphics[scale=0.7]{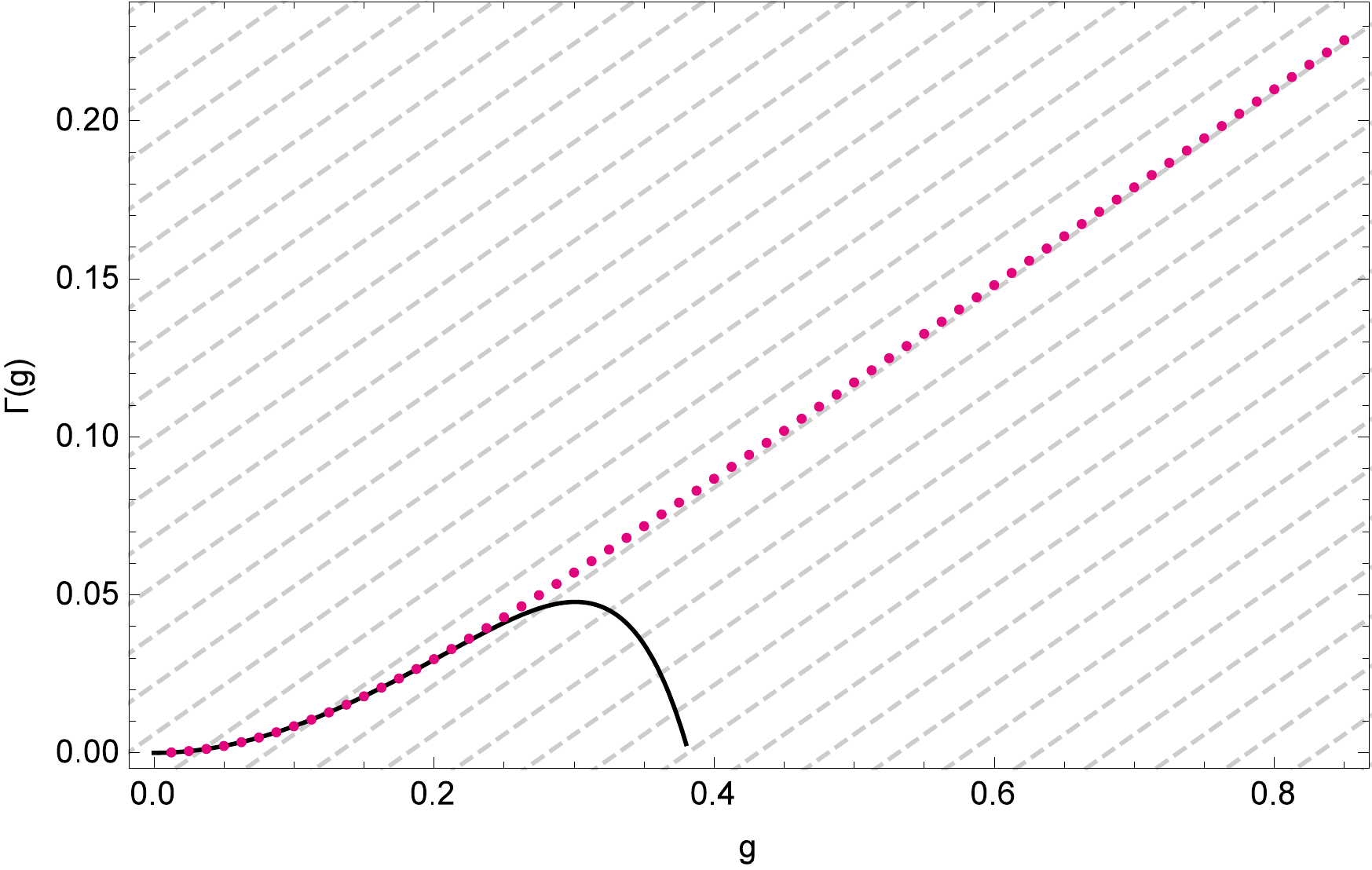}
\end{tabular}
\caption{Numerically evaluated cusp anomalous dimension $\Gc$
for $\phi=\pi/4,\;\theta=4\pi/10$ in a wide range of the coupling $g$.
Solid line shows the $4$-loop
perturbation theory prediction of \cite{Makeenko:2006ds,Drukker:2011za,Correa:2012nk,Henn:2013wfa}.
Dashed lines indicate the leading strong coupling prediction
for the slope of the function at $g\to\infty$.
}
}

Let us now present the numerical results we obtained. First, we have evaluated $\Gc$ for a wide range of the coupling from $g=0$ up to $g=0.85$ at fixed values of the angles $\phi={\pi}/{4},\;\theta={4\pi}/{10}$. The results are given in Table \ref{numtab}.
A fit of our data at weak coupling gives
\beqa
&&\Gc\(\phi=\frac{\pi}{4},\theta=\frac{4\pi}{10},g\)\simeq
  0.8843331608401797458041129816\; g^2\\
\nn&&-4.7002219374112776568286369\; g^4+
  37.481607207831059124394\;  g^6\\
\nn&&-
 321.37797809257617613\; g^8+
 2845.9019611906881\; g^{10}\\
\nn&&-
25984.505154213\; g^{12}+{\cal O}(g^{14})
\eeqa
which agrees with the analytical perturbative
result of \cite{Makeenko:2006ds,Drukker:2011za,Correa:2012nk,Henn:2013wfa} with $10^{-29}g^2+10^{-25}g^4+10^{-21}g^6+10^{-18}g^8$ error. The terms $g^{10}$ and $g^{12}$ above also give a numerical prediction for the five- and six-loop coefficients. One could try to get an analytic prediction for them by fitting the numerical data as a combination of some basis harmonic polylogarithms. This would require higher precision of course but should be possible to do (e.g. in \cite{Gromov:2015vua} more than 60 digits of precision were reached).

At strong coupling only the leading classical result is known in explicit form at generic angles.
It can be extracted from \cite{Drukker:2011za,Gromov:2012eu} which
gives the $\sim g$ coefficient. For $\phi=\frac{\pi}{4}$ and $\theta=\frac{4\pi}{10}$
it gives $\Gamma^{\rm classical}_{\rm cusp}\simeq 0.3122881 g$. Fitting our data we get
\beq
\Gc\(\phi=\frac{\pi}4,\theta=\frac{4\pi}{10},g\) \simeq 0.3122892\;g
-0.0410591+\frac{
 0.00073853}{g}+
{\cal O}\left(\frac{1}{g^2}\right)
\eeq
which agrees nicely with
the
AdS/CFT prediction. Let us mention that at strong coupling it requires some effort to get high precision since we need to keep many terms in the expansion \eq{fgexp}. It would be interesting to compare our result for the $g^0$ term with the 1-loop prediction of \cite{Drukker:2011za} which is written in an implicit form. One should also be able to derive the one-loop correction in a simpler and more general way by using the algebraic curve as in \cite{Gromov:2007aq}.  On Fig. \ref{fign} one can see that our data clearly interpolates between gauge and string theory results. 

\FIGURE[ht]{
\label{3dplot}
\begin{tabular}{cc}
\includegraphics[natwidth=960,natheight=720,scale=0.6]{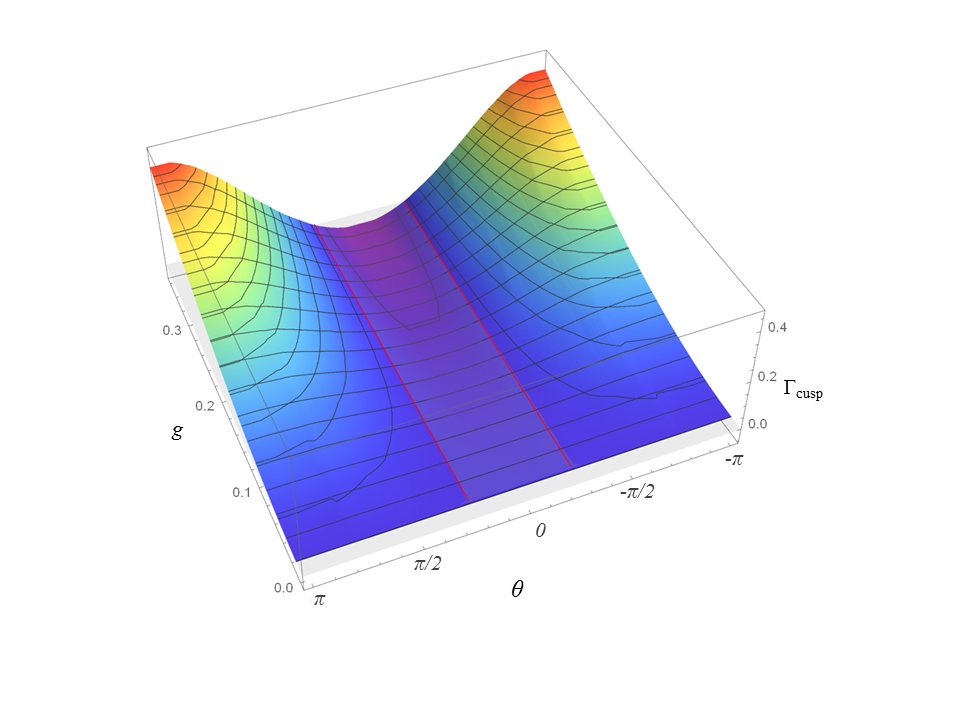}\\
\end{tabular}
\caption{{A 3d plot of $\Gc$ at fixed $\phi=\pi/4$ in a range of values of the coupling $g$ and the angle $\theta$, generated from $\sim 800$ data points. We also added a semi-transparent purple plane located at $\Gc=0$, and two red lines corresponding to the BPS configuration $\theta=\pm\phi$ for which $\Gc=0$ (i.e. $\theta=\pm\pi/4$ in our case).} }
}
In addition, on Fig. \ref{3dplot} we show our numerical data for the generalized cusp anomalous dimension at $\phi=\pi/4$ for various values of $\theta$ and of the coupling. One can clearly see in particular the straight lines corresponding to the BPS regime $\phi=\theta$ when $\Gc$ is zero. We covered the full range of $\theta$ from $-\pi$ to $\pi$, and on the plot one can see that as expected $\Gc$ is a smooth and $2\pi$-periodic function of this angle, invariant under $\theta\to-\theta$.

\section{Weak coupling solution}

In section \ref{sec:nearbps} we constructed the solution of the QSC in the near-BPS limit $\phi-\theta\to0$. In this section we will describe the solution for \textit{arbitrary} angles, at leading order in $g$. We will discuss the case $L=0$.

At weak coupling the cuts degenerate into poles, but the singular part is typically suppressed by the coupling so one could expect $\bP_a$ to be regular at leading order. However the asymptotics \eq{cuspas} mean that we have to allow a $1/\sqrt{u}$ singularity in $\bP_1,\bP_2$. This leads to the ansatz
\beqa
	\bP_1&=&C_1\frac{e^{\theta  u}}{\sqrt{u}},\ \ \ \ \
	\bP_2=C_2\frac{e^{-\theta  u
   }}{\sqrt{u}},
	\\ \nn
	\bP_3&=&
e^{\theta u}(C_3u^{3/2}+C_4u^{1/2}),
\ \ \ \ \ \
	 \bP_4=e^{-\theta u}(C_5u^{3/2}+C_6u^{1/2})\;.
\eeqa
Then all the coefficients are completely fixed by asymptotics (up to a rescaling \eq{Presc}), giving
\beqa
	\bP_1&=&\sqrt{\eps}\frac{e^{\theta  u}}{\sqrt{u}},\ \ \ \ \
	\bP_2=\sqrt{\eps}\frac{e^{-\theta  u
   }}{\sqrt{u}},
	\\ \nn
	\bP_3&=&
 \eps^{3/2}u^{3/2}e^{\theta u}(1+b/u),
\ \ \ \ \ \
	 \bP_4=-\eps^{3/2}u^{3/2}e^{-\theta u}(1-b/u)
\eeqa
where
\beq
	b=\frac{2 \cos \theta  \cos \phi+\cos 2 \theta -3}{2 \sin\theta (\cos \theta -\cos \phi)}
\eeq
and $\eps$ is defined in \eq{epsab}.

Let us now discuss $\mu_{ab}$. At leading order in the weak coupling expansion we expect that in the general expression
\beq
\label{muom3}
	\mu_{ab}=\frac{1}{2}Q^-_{ab|ij}\omega^{ij}
\eeq
only $\omega^{12}$ will contribute, in analogy with the undeformed QSC \cite{PmuPRL,PmuLong,Marboe:2014gma} as this also what happens in the asymptotic large $L$ regime. Based on our large $u$ prescription $\omega^{12}\sim e^{2\pi |u|}$ and the near-BPS solution \eq{muLO}, it is natural to take
\beq
	\omega^{12}=\const\cdot\sinh(2\pi u)\;.
\eeq
In fact, for computing higher orders in the weak coupling expansion it should be better to completely avoid calculating $\omega^{ij}$ and apply instead the equations \eq{qg1}, \eq{qg2} we used in the numerics.
For the functions $Q_{ab|12}$ we can make an ansatz as polynomials whose degree is determined by the asymptotics of $Q_{ab|12}$, times $e^{\pm2\theta u}$ in accordance with asymptotics again. Also, we expect that those of the functions $Q_{ab|12}$ which do not have exponential asymptotics should be either even or odd. Thus we use the following ansatz:
\beq
	\{\mu_{12}^+,\mu_{13}^+,\mu_{14}^+,\mu_{24}^+,\mu_{34}^+\}=
\eeq
\beqa
\label{muansw}
	\nn &&
	\sinh(2\pi u)\left\{
	D_1,\
	e^{2 \theta  u}(D_2+uD_3),\
	D_4u^2+D_5,\
	\right.
	\\
	&&\left.\nn
	e^{-2
   \theta  u} (D_6+uD_7),\
	D_8u^4+D_9u^2+D_{10}
   \right\}\ .
\eeqa
To fix the constants $D_i$ appearing here we use the difference equation on $\mu_{ab}$ following from the $\bP\mu$-system equations \eq{tPd}, \eq{tmud}:
\beq
\label{mudif}
	\mu_{ab}^{++}-\mu_{ab}=\mu_{ac}\bP^c\bP_b-\mu_{bc}\bP^c\bP_a
\eeq
where $\bP^a$ are related to $\bP_a$ by \eq{Pud}.
This equation fixes all the constants except one, and we get
\beq \nn
	\{\mu_{12}^+,\mu_{13}^+,\mu_{14}^+,\mu_{24}^+,\mu_{34}^+\}=
\eeq
\beqa
\label{muweak}
	\nn &&
	{R \sinh(2\pi u)}\left\{-\frac{\sin \theta }{\epsilon },
	\frac{e^{2 \theta  u}}{2}  (2 u-\cot \theta
   ),
	\frac{\sin \theta}{4}   \left(-\frac{2}{ \sin ^2\theta} +4 u^2+1\right),
	\right.
	\\
	&&\left.
	-\frac{1}{2} e^{-2
   \theta  u} (\cot \theta +2 u),
	\frac{1}{16} \left(4 u^2+1\right)^2 \epsilon  \sin \theta
   \right\}\ .
\eeqa
Going to higher orders in $g$ (see below) we also found that the constant $R$ and $\omega^{12}$ scale as $\sim 1/g^2$.

The $\bQ$-functions can be found from the 4th order Baxter equation on $\bQ_i$ with coefficients built from $\bP_a$ (see \cite{Alfimov:2014bwa} for its derivation). They turn out to be written in terms of generalized $\eta$-functions defined as
\beq
\label{defetam}
\eta_{s_1,\dots,s_k}^{z_1,\dots,z_k}(u)\equiv \sum_{n_1 > n_2 > \dots > n_k \geq 0}\frac{z_1^{n_1}\dots z_k^{n_k}}{(u+i n_1)^{s_1}\dots (u+i n_k)^{s_k}}
\eeq
For the case when all twists $z_i$ are equal to 1 such functions already appeared in the weak coupling calculations of \cite{Leurent:2013mr,Marboe:2014gma}. Importantly, all operations needed for the iterative procedure of \cite{Gromov:2015vua} (e.g. expressing the product as a linear combination or solving equations of the kind $f(u+i)-f(u)=\eta^{z_1,\dots,z_k}_{s_1,\dots,s_k}(u)$) can be carried out for these functions as we describe in Appendix \ref{sec:eta}. For future applications, we attach to this submission a Mathematica notebook implementing some of these operations on the generalized $\eta$-functions.

In terms of $\eta$-functions we found the following four linearly independent solutions of the fourth order Baxter equation:
\beqa
	\bQ_1&=&\sqrt{u} e^{u \phi },
	\\ \nn
	\bQ_2&=&\sqrt{u} e^{-u \phi },
	\\ \nn
	\bQ_3&=&\frac{e^{u \phi } (\sin \phi +i u
   (\eta_1^{z}-\eta_1^1) (\cos \theta -\cos \phi ))}{\sqrt{u} (\cos \phi -\cos \theta
   )},
	\\ \nn
	\bQ_4&=&\frac{e^{-u \phi } (-\sin \phi +i u (\eta_1^{\bar z}-\eta_1^1) (\cos\theta -\cos \phi
   ))}{\sqrt{u} (\cos \phi -\cos \theta )}
\eeqa
where $z=e^{2i\phi},\bar z=e^{-2i\phi}$. The true $\bQ$-functions should be identified with appropriate linear combinations of these four solutions.

To fix the anomalous dimension $\Gc$ one needs to go to higher orders in $g$. This can be done using the iterative algorithm of \cite{Gromov:2015vua}
for which $\bP_a$ and $\bQ_i$ we have found serve as a starting point. Notice that the weak coupling algorithm of \cite{Marboe:2014gma} is not directly applicable in our situation, as all $\bP_a$ are of the same order $\sim g^0$ and none of them are small at weak coupling. In particular, none of the five independent equations among \eq{mudif} decouple from the rest at leading order. However the universal iterative method of \cite{Gromov:2015vua} works well, and we used it to compute the $\bP$- and $\bQ$-functions at higher orders\footnote{To simplify intermediate expressions we used several Mathematica packages \cite{packages}}. In particular we reproduced the one-loop prediction
\beq
	\Gc=2g^2\frac{\cos\phi-\cos\theta}{\sin\phi}\phi+\cO(g^4)
\eeq
directly at any $\phi$ and $\theta$. The details of this calculation will be presented elsewhere. Using this method it is certainly possible to also reach much higher loops.

\section{Conclusions}

In this paper we present the modifications needed in the Quantum Spectral Curve to describe the generalized cusp anomalous dimension.
We show that the main new ingredient of the boundary TBA formulation -- the boundary reflection phase \cite{Correa:2012hh,Drukker:2012de}  -- is mapped to a simple modification of the $\omega^{12}$ asymptotics.
Our proposal is consistent with the known near-BPS solution, and we also computed the subleading term in the near-BPS expansion at any coupling.
The result matches perfectly the known perturbative predictions, providing a deep test of the QSC for this model. As supplementary material to this arXiv submission we added a Mathematica notebook with the perturbative expansion of our all-loop result. We also attach a notebook which should be useful for a perturbative solution of the QSC at generic angles, and a file with numerical data at finite coupling having $\sim 20$ digits precision.

Curiously,
our modification of the asymptotics for
the component $\omega^{12}$ of the periodic anti-symmetric matrix $\omega^{ij}$
is very similar to that needed for the analytic continuation in Lorentz spin for the twist-2 local
operators where the $\omega^{13}$ asymptotics was relaxed to be exponentially large \cite{Gromov:2014bva,Alfimov:2014bwa,Gromov:2015wca}.
It seems to be a common feature of non-local operators.
It would be interesting to classify all consistent asymptotics of this kind and find the corresponding integrable observables.

Our results together with \cite{DimaKazTwist} elucidate the structure of the QSC for
models with twisted boundary conditions. Extension to the $q$-deformation seems also to be straightforward.
Generalization of the QSC for other boundary problems such as \cite{Bajnok:2013wsa} should help to understand some of their still mysterious features.

The drastic simplification of the TBA we have achieved calls for a systematic exploration of $\Gc$ in various regimes, with the hope of revealing new structures. One should now be able to reach much higher loop orders in the perturbative expansion with arbitrary $\phi,\theta$ using the methods of \cite{Gromov:2015vua,Marboe:2014gma,Marboe:2014sya},
 study analytically various special cases such as the ``ladders'' limit \cite{Correa:2012nk,ladders}, and try to extend the link to matrix models observed in \cite{Gromov:2012eu,Gromov:2013qga,Sizov:2013joa}.
It would be also interesting to explore the connection to the supersymmetric hydrogenlike bound states of massive W-bosons in $\cN=4$ SYM \cite{Caron-Huot:2014gia}.

While a numerical solution of the TBA is additionally complicated by
the infinite sums which diverge for real $\phi$ and $\theta$ \cite{Bajnok:2013sya},
the simple high-precision numerical method of \cite{Gromov:2015wca} for the QSC is applicable almost directly.
Computing $\Gc$ numerically in a wide range of the coupling we found perfect interpolation between gauge theory and string theory predictions.
It is of course also interesting to develop a systematic analytical method for the strong coupling expansion and extend the celebrated string theory predictions \cite{Forini:2010ek,Drukker:2011za}.

\section*{Acknowledgements}
We thank M. Alfimov, S. Caron-Huot, A. Cavaglia, N. Drukker, V. Kazakov, A. Sever, G. Sizov, R. Tateo, S. Valatka and D. Volin for discussions. We are especially grateful to M. Alfimov and A. Cavaglia for collaboration at the early stages of this project. We are also grateful to V. Kazakov, S. Leurent and D. Volin for sharing their draft of \cite{DimaKazTwist} before publication. The research leading to these results has received funding from the People Programme
(Marie Curie Actions) of the European Union's Seventh Framework Programme FP7/2007-
2013/ under REA Grant Agreement No 317089 (GATIS).
 We wish to thank
SFTC for support from Consolidated
grant number ST/J002798/1.
N.G. would like to thank FAPESP grant 2011/11973-4 for funding his visit to ICTP-SAIFR during
January 2015 where part of this work was done.

\appendix

\section{The anomalous dimension from asymptotics}
\label{sec:asympDL}

Her we present the explicit expression we got for the conformal dimension $\Delta$ in terms of the coefficients $a_i,b_i$ in the large $u$ expansion of the $\bP$-functions (see Eq. \eq{cuspas}), for any $L$. It reads
\beqa\label{DLfull}
\nn
	\Delta^2&=&
	-a_1
	\left[\frac{a_2   (\cos \theta -\cos \phi
   )^3}{(L+1)\sin\theta\sin^2\phi}-
	\frac{b_2  (\cos \theta -\cos \phi
   )^3}{(L+1)\sin\theta\sin^2\phi}
	+F(\theta,\phi,L)
	\right]
	\\ \nn
	&-& \frac{ a_1^2 (\cos \theta  \cos \phi -1) (\cos \theta -\cos \phi )^2}{\sin^2\theta\sin^2\phi}
	+\frac{a_3 (\cos \theta -\cos \phi )^3}{(L+1)\sin\theta\sin^2\phi}
	\\
	&-&\frac{a_2  (\cos \theta -\cos \phi )^2 (-2 \cos \theta  \cos
   \phi +(L+1) \cos 2 \theta -L+1)}{2 (L+1)\sin^2\theta\sin^2\phi}
	\\ \nn
	&-&
	\frac{b_3
    (\cos \theta -\cos \phi )^3}{(L+1)\sin\theta\sin^2\phi}+\frac{b_2 L
   (\cos \theta  \cos \phi -1) (\cos \theta -\cos \phi )^2}{(L+1)\sin^2\theta\sin^2\phi}
	\\ \nn
	&+&\frac{(2 L+1) L}{24\sin^2\theta\sin^2\phi}
  \[	\frac{\mathstrut}{\mathstrut}{\cos \theta  \(\cos 3 \phi -10 \cos
   \phi \)} + \cos 3 \theta  \cos \phi +8
   \]
	-\frac{L(1-L)}{3}
\eeqa
where

\beqa
\nn
	F(\theta,\phi,L)&=&\frac{ (\cos \theta -\cos \phi )}{4\sin^3\theta\sin^2\phi}
	\[-2 (5
   L+4) \cos \theta  \cos \phi
		+(L+2) \cos 2 \phi +7 L+4
	\frac{\mathstrut}{\mathstrut}
	\right.
	\\
		&+&\left.
		\cos 2 \theta  \(2 L \cos \theta  \cos \phi
   +L \cos 2 \phi -L+2\)
	\frac{\mathstrut}{\mathstrut}
	\]
\eeqa

\section{Asymptotics of $\bQ$-functions}
\label{sec:asympQ}

Similarly to the asymptotics of $\bP_a$ given in \eq{cuspas} in the main text, we found that the asymptotics of $\bQ_i$ have the form (with $C$ an arbitrary constant)
\beqa
\label{Qfullas}
\bQ_1(u)&\simeq&C\epsilon'^{1/2}\;u^{1/2+\Delta}\; e^{+\phi u} F(+u)\;\;,\;\;F(u)=1+c_1/u+c_2/u^2+c_3/u^3+\dots\\
\nn
\bQ_2(u)&\simeq&C\epsilon'^{1/2}\;u^{1/2+\Delta}\; e^{-\phi u} F(-u)\\
\nn
\bQ_3(u)&\simeq&\frac{1}{C}\epsilon'^{3/2}\;u^{1/2-\Delta}\; e^{+\phi u} G(+u)\;\;,\;\;G(u)=1+d_1/u+d_2/u^2+d_3/u^3+\dots\\
\nn
\bQ_4(u)&\simeq&-\frac{1}{C}\epsilon'^{3/2}\;u^{1/2-\Delta}\; e^{-\phi u} G(-u)
\eeqa
while $\bQ$'s with upper and lower indices are related as in \eq{Pud},
\beq
	\bQ^1=-\bQ_4,\ \bQ^2=+\bQ_3,\ \bQ^3=-\bQ_2,\ \bQ^4=+\bQ_1\
\eeq
The coefficients are constrained by
\beq
\epsilon'^2=-\frac{i (\cos \theta -\cos \phi)^2}{2 \Delta \sin^2 \phi}\;\;,\;\;
c_1-d_1=-\frac{\Delta (2 \cos \theta  \cos \phi+\cos 2 \phi -3)}{2 \sin\phi (\cos \theta -\cos \phi)}
\eeq
While $\Delta$ enters the powers in the asymptotics of $\bQ_i$, the remaining conserved charge $L$ is encoded in the large $u$ expansion coefficients as
\beqa
	L(L+2)&&=c_2 \left[\frac{d_1 \csc ^2\theta  \csc \phi  (\cos \phi -\cos \theta
   )^3}{\Delta }
	\right. \\ \nn && \left.
	+\frac{(\Delta -1) \csc ^2\theta  \csc ^2\phi  (\cos \theta
   \cos \phi -1) (\cos \phi -\cos \theta )^2}{\Delta }\right]
	\\ \nn &&
	+\frac{c_3 \csc
   ^2\theta  \csc \phi  (\cos \theta -\cos \phi )^3}{\Delta }
	+\frac{d_3 \csc
   ^2\theta  \csc \phi  (\cos \phi -\cos \theta )^3}{\Delta }
	\\ \nn &&
	+d_1
   \left[\frac{d_2 \csc ^2\theta  \csc \phi  (\cos \theta -\cos \phi
   )^3}{\Delta }+F_1(\theta,\phi,\Delta)
	\right]
	\\ \nn &&
	+\frac{d_2 \csc ^2\theta  \csc ^2\phi  (\cos \phi
   -\cos \theta )^2 \left(\Delta  \sin ^2\phi +\cos \theta  \cos \phi
   -1\right)}{\Delta }
	\\ \nn &&
	-d_1^2 \csc ^2\theta  \csc ^2\phi  (\cos \phi -\cos
   \theta )^2 (\cos \theta  \cos \phi -1)
	\\ \nn &&
	+\frac{1}{24} \left[-(\Delta -1) (2
   \Delta -1) (\cos \theta -2) \cot ^2\left(\frac{\theta }{2}\right) \sec
   ^2\left(\frac{\phi }{2}\right)
			\right. \\ \nn && \left.
	-4 (\Delta -1) (2 \Delta -1) \cot \theta  \csc
   \theta  \cos \phi
			\right. \\ \nn && \left.
	+(\Delta -1) (2 \Delta -1) (\cos \theta +2) \tan
   ^2\left(\frac{\theta }{2}\right) \csc ^2\left(\frac{\phi }{2}\right)	
	+8 ((\Delta -3)
   \Delta -1)
\right]
\eeqa
where we denote
\beq
	\csc\theta\equiv1/\sin\theta,\ \ \sec\theta\equiv1/\cos\theta
\eeq
and
\beqa
\nn
	F_1(\theta,\phi,\Delta)&&=\frac{1}{4} \csc ^2\theta  \csc ^3\phi  (\cos \theta -\cos
   \phi ) \[2 \cos \theta  \cos \phi  ((\Delta -1) \cos 2 \phi -5 \Delta
   +1)
		\right. \\  && \left.
	+\cos 2 \theta  ((\Delta -1) \cos 2 \phi +\Delta +1)-(\Delta -3) \cos 2
   \phi +7 \Delta -3\]
\eeqa

\section{The leading near-BPS solution at any $L$}
\label{sec:appbps}

Let us present explicitly the leading order near-BPS solution of the $\bP\mu$ system at any $L$. It was constructed in \cite{Gromov:2013qga} and below we write it in our conventions. Most importantly, imposing the asymptotics \eq{cuspas} and \eq{epsab} we recovered from \eq{deltaas} the all-loop results of \cite{Gromov:2013qga} for the near-BPS cusp anomalous dimension at nonzero $L$, providing a stringent test of the asymptotics we propose in this paper\footnote{We checked the matching explicitly for the first several $L$'s}.

The solution has the following form. First, the components of $\mu_{ab}$ are
\beq
\label{muanyL}
	\mu_{12}^{(0)}=A\sinh(2\pi u),\ \mu_{13}^{(0)}=(-1)^L,\ \mu_{14}^{(0)}=0,\
	\mu_{24}^{(0)}=(-1)^{L+1},\ \mu_{34}^{(0)}=0
\eeq
Second, the $\bP$-functions read
\beqa
	\bP_1^{(0)}&=&K\sqrt{A}\sqrt{u}e^{\theta u}\frac{\tilde F(x)}{x^{L+1}}\ ,\\ \nn
	\bP_2^{(0)}&=&K\sqrt{A}\sqrt{u}e^{-\theta u}\frac{\tilde F(-x)}{x^{L+1}}\ ,\\ \nn
	\bP_3^{(0)}&=&\frac{K}{\sqrt{A}}\sqrt{u}e^{g\theta(x-1/x)}P_L(x)\ ,\\ \nn
	\bP_4^{(0)}&=&(-1)^{L}\frac{K}{\sqrt{A}}\sqrt{u}e^{-g\theta(x-1/x)}P_L(-x)\ .
\eeqa
Here $A$ is a constant which can be set to 1 via a rescaling \eq{Presc}, \eq{muresc} while the constant $K\sim\sqrt{\theta-\phi}$ can be fixed from asymptotics \eq{cuspas}, \eq{epsab}. The function $F(x)$ is a power series
\beq
	F(x)=1+\sum_{n=1}^\infty f_n x^n\ ,
\eeq
which satisfies
\beq
	e^{2g\theta x}x^{L+1}F(x)+(-1)^Le^{-2g\theta/x}{x^{L+1}}\tilde F(-x)
	=\sinh(2\pi u)e^{2g\theta(x-1/x)}P_L(x)
\eeq
and is fixed as
\beq
	F(x)=e^{-2g\theta x}x^{-L-1}\[\sinh(2\pi u)e^{2g\theta(x-1/x)}P_L(x)\]_+
\eeq
where $[f]_+$ denotes the part of the Laurent expansion of $f(x)$ with positive powers of $x$. Finally, the Laurent polynomial $P_L(x)$ reads
\beq
	P_L(x)=\frac{1}{\det {\cal M}_{2L}}\left|\begin{matrix}
	I_1^{\theta}& I_0^{\theta}& \cdots & I_{2-2L}^{\theta}  &I_{1-2L}^{\theta}\\
	I_2^{\theta}& I_1^{\theta}& \cdots & I_{3-2L}^{\theta} &I_{2-2L}^{\theta}\\
	\vdots      &  \vdots     &\ddots & \vdots            &\vdots           \\
	I_{2L}^{\theta}& I_{2L-1}^{\theta}& \cdots & I_{1}^{\theta}  &I_{0}^{\theta}\\
	x^{-L}& x^{1-L}& \cdots & x^{L-1} &x^{L}\\
	\end{matrix}\right|
\eeq
where
\beq
	{\cal M}_{N}=\begin{pmatrix}
	I_1^{\theta}& I_0^{\theta}& \cdots & I_{2-N}^{\theta}  &I_{1-N}^{\theta}\\
	I_2^{\theta}& I_1^{\theta}& \cdots & I_{3-N}^{\theta} &I_{2-N}^{\theta}\\
	\vdots      &  \vdots     &\ddots & \vdots            &\vdots           \\
	I_{N}^{\theta}& I_{N-1}^{\theta}& \cdots & I_{1}^{\theta}  &I_{0}^{\theta}\\
	I_{N+1}^{\theta}& I_{N}^{\theta}& \cdots & I_{2}^{\theta} &I_{1}^{\theta}
	\end{pmatrix}.
\eeq
Notice also that
\beq
	P_L(1/x)=P_L(-x)
\eeq
From this solution using \eq{deltaas} we recover the result of \cite{Gromov:2013qga} for the cusp anomalous dimension,
\beq
	\Gc=L+\frac{\phi-\theta}{4}\partial_\theta\log\frac{\det{\cal M}_{2L+1}}{\det{\cal M}_{2L-1}}
	+\cO((\phi-\theta)^2)\ .
\eeq

\section{The near-BPS result for small $\phi$}
\label{sec:phi0}

Here we present our result for the cusp anomalous dimension in the near-BPS limit, further expanded for $\phi\to 0$. Let us remind that in our notation
\beq
	\Delta=\frac{\cos\phi-\cos\theta}{\sin\phi}\Delta^{(1)}(\phi)
	+\(\frac{\cos\phi-\cos\theta}{\sin\phi}\)^2\Delta^{(2)}(\phi)+\cO((\phi-\theta)^3)
\eeq
Carefully taking the $\phi\to 0$ limit in our explicit result \eq{D2res} we found
\beqa
	\Delta^{(2)}(\phi)=f_2(g)\phi^2+\cO(\phi^3)
\eeqa
with
\beqa\nn
	f_2(g)&=&-\frac{4g^2}{\pi^2}\(\frac{I_2(4\pi g)}{I_1(4\pi g)}\)^2
	-\frac{1}{2}\oint\frac{du_x}{2\pi i}\oint\frac{du_y}{2\pi i}
	\frac{1}{4\pi i}\Gamma_0(u_x-u_y)F(x,y)\\
\eeqa
where the integrals go around the cut $[-2g,2g]$, the kernel $\Gamma_0$ is defined in \eq{Gamma0}, and most importantly
\beqa
	F(x,y)&&=-\frac{8 i \sinh \left(2 \pi  u_x\right) u_x  u_y x^2S_0(y) }{I_1(4 g \pi ){}^2}
	\\ \nn
	&&+S_0(y){}^2\left[\frac{8 i x y I_2(4 g \pi ) u_x u_y}{g
   \pi  \left(x^2-1\right) I_1(4 g \pi ){}^3}-\frac{8 i x y I_2(4 g \pi ) u_x u_y}{g \pi
   \left(y^2-1\right) I_1(4 g \pi ){}^3}+\frac{32 i x y u_x u_y}{I_1(4 g \pi ){}^2}\right]
	\\ \nn
	&&+\sinh ^2\left(2 \pi  u_y\right) \left[\frac{4 i x y I_2(4 g \pi ) u_x u_y}{g
   \pi  \left(x^2-1\right) I_1(4 g \pi ){}^3}+\frac{16 i x y u_x u_y}{I_1(4 g \pi
   ){}^2}\right]
	\\ \nn
	&&+\sinh \left(2 \pi  u_y\right) \left[\frac{4 i x u_x u_y
   y^2}{\left(x^2-1\right) I_1(4 g \pi )}
	-\frac{8 i x \sinh \left(2 \pi  u_x\right) u_x u_y
   y}{I_1(4 g \pi ){}^2}
	-\frac{8 i u_x u_y S_1(x) y}{g I_1(4 g \pi ){}^2}
	\right.
	\\ \nn
	&& \left.
	-\frac{16 i x u_x
   u_y}{\left(y^2-1\right) I_1(4 g \pi )}+\left(-\frac{8 i x y I_2(4 g \pi ) u_x u_y}{g \pi
   \left(x^2-1\right) I_1(4 g \pi ){}^3}-\frac{32 i x y u_x u_y}{I_1(4 g \pi ){}^2}\right)
   S_0(y)\right]
	\\ \nn
	&&+S_1(y)\left[\frac{8 i x y u_x u_y}{g \left(x^2-1\right) I_1(4 g \pi )}-\frac{8 i x
   y u_x u_y}{g \left(y^2-1\right) I_1(4 g \pi )}\right]
	\\ \nn
	&&
	+S_0(x) \left[S_0(y)\left(\frac{16 i
   u_x u_y}{I_1(4 g \pi ){}^2}-\frac{16 i y^2 u_x u_y}{I_1(4 g \pi ){}^2}\right)
   -\frac{4 i x I_2(4 g \pi ) u_x u_y S_1(y)}{g^2 \pi  \left(x^2-1\right) I_1(4 g \pi
   ){}^3}\right]
	\\ \nn
	&+&S_0(y) \left[\frac{8 i x u_x u_y y^2}{\left(x^2-1\right) I_1(4 g \pi
   )}
	+\frac{8 i x u_x u_y}{I_1(4 g \pi )}
	-\frac{8 i x u_x u_y}{\left(x^2-1\right) I_1(4 g \pi
   )}
	+\frac{32 i x u_x u_y}{\left(y^2-1\right) I_1(4 g \pi )}
	\right. \\ \nn
	&&\left. +S_1(x)\left(-\frac{4 i x I_2(4 g \pi
   ) u_x u_y}{g^2 \pi  \left(x^2-1\right) I_1(4 g \pi ){}^3}-\frac{16 i x u_x u_y}{g I_1(4 g
   \pi ){}^2}\right)
	\right. \\ \nn
	&& \left.
	+S_1(y)\left(\frac{4 i x I_2(4 g \pi ) u_x u_y}{g^2 \pi
   \left(x^2-1\right) I_1(4 g \pi ){}^3}+\frac{16 i x u_x u_y}{g I_1(4 g \pi ){}^2}\right)
   \right]
\eeqa
Here we used the notation
\beq
	S_0(x)=\sum\limits_{n=1}^\infty I_{2n+1}(4\pi g)/x^{2n+1},\ \ \
	S_1(x)=\sum\limits_{n=1}^\infty \frac{2nI_{2n}(4\pi g)}{\pi x^{2n}}\textbf{}
\eeq

\newpage

\section{Weak coupling predictions at five and six loops}
\label{sec:weak5}

From our all-loop result \eq{D2res} it is straightforward to obtain a prediction for a part of the full anomalous dimension at five and six loops, namely for the coefficients $\gamma_5^{(2)}(\phi)$ and $\gamma_6^{(2)}(\phi)$ in
\eq{dstruc}. We found them to be
\beqa
\label{our5loop}
\gamma_5^{(2)}(\phi)&&=
	3360
	\[
	\text{Li}_9(e^{-2 i \phi})
	+\text{Li}_9(e^{2 i \phi })
	\]
		-2156 i \phi
	\[
	\text{Li}_8(e^{2 i \phi })-\text{Li}_8(e^{-2 i \phi })
	\]
		\\ \nn
	&&
	-8 \left(62 \phi ^2+15 \pi ^2\right) \[
	\text{Li}_7(e^{2 i \phi })+\text{Li}_7(e^{-2 i \phi })
	\]
		\\ \nn &&
+\frac{20}{3} i
   \left(49 \pi ^2 \phi -29 \phi ^3\right) \[\text{Li}_6(e^{2 i \phi })-\text{Li}_6(e^{-2 i \phi })\]
	\\ \nn &&
	-\frac{8}{3} \left(73 \phi ^4-87 \pi ^2 \phi ^2+6
   \pi ^4\right)
	\[
	\text{Li}_5(e^{2 i \phi })+\text{Li}_5(e^{-2 i \phi })
	\]
	\\ \nn &&
	+\frac{4}{3} i
   \left(65 \phi ^5-94 \pi ^2 \phi ^3+29 \pi ^4 \phi \right)
	\[
	\text{Li}_4(e^{2 i \phi})-\text{Li}_4(e^{-2 i \phi})
	\]
		\\ \nn &&
	-\frac{8}{9} (\pi -\phi ) (\phi +\pi ) \left(33 \phi ^4-31 \pi ^2 \phi ^2+2 \pi ^4\right)
	\[
   \text{Li}_3(e^{2 i \phi })+\text{Li}_3(e^{-2 i \phi })
	\]
	\\ \nn
	&&+\frac{32}{45} i \phi  \left(7 \pi ^2-12 \phi ^2\right) \left(\pi ^2-\phi ^2\right)^2
	\[
   \text{Li}_2(e^{2 i \phi })-\text{Li}_2(e^{-2 i \phi })
	\]
	\\ \nn &&
	+\frac{32}{5} \phi ^2 \left(\phi ^2-\pi ^2\right)^3 \[\log (1-e^{2 i \phi })+\log (1-e^{-2 i \phi })\]
	\\ \nn &&
	+\frac{16}{45} \left[83 \zeta (3) \phi ^6-15 \left(8
   \pi ^2 \zeta (3)+31 \zeta (5)\right) \phi ^4+3 \left(9 \pi ^4 \zeta (3)+85 \pi ^2 \zeta (5)+930 \zeta
   (7)\right) \phi ^2
	\right. \\ \nn && \left.
	-18900 \zeta (9)+675 \pi ^2 \zeta (7)+90 \pi ^4 \zeta (5)+10 \pi ^6 \zeta
   (3)\right]
\eeqa

\newpage

and
\beqa
\nn
	\gamma_6^{(2)}(\phi)&&=
	-41580 \left[\text{Li}_{11}(e^{-2 i \phi })+\text{Li}_{11}(e^{2 i \phi
   })\right]
	+25704
   i \phi  \left[\text{Li}_{10}(e^{2 i \phi })-\text{Li}_{10}(e^{-2 i \phi
   })\right]
	\\
	&&
	+168 (35 \phi ^2+12 \pi ^2)
   \left[\text{Li}_9(e^{-2 i \phi })+\text{Li}_9(e^{2 i \phi })\right]
	\\ \nn &&
	-\frac{56}{3} i (241 \pi ^2 \phi -137 \phi ^3)
	\left[\text{Li}_8(e^{2 i \phi
   })-\text{Li}_8(e^{-2 i \phi })\right]
			\\ \nn &&
	+\frac{8}{3} \left(943 \phi ^4-1150 \pi ^2 \phi
   ^2+91 \pi ^4\right)
	\left[\text{Li}_7(e^{-2 i \phi })+\text{Li}_7(e^{2 i \phi
   })\right]
		\\ \nn &&
	-\frac{4}{9} i
   \left(2661 \phi ^5-3754 \pi ^2 \phi ^3+1077 \pi ^4 \phi \right)
	\left[\text{Li}_6(e^{2 i \phi
   })-\text{Li}_6(e^{-2 i \phi })\right]
				\\ \nn &&
	+\frac{8}{45} \left(-2299 \phi ^6+3970 \pi ^2 \phi ^4-1835 \pi ^4 \phi ^2+148 \pi
   ^6\right)
	\left[\text{Li}_5(e^{-2 i \phi })+\text{Li}_5(e^{2 i \phi })\right]
		\\ \nn &&
	-\frac{16}{45} i (\pi -\phi ) \phi  (\phi +\pi )
   \left(351 \phi ^4-449 \pi ^2 \phi ^2+154 \pi ^4\right)
	\left[\text{Li}_4(e^{2 i \phi
   })-\text{Li}_4(e^{-2 i \phi })\right]
		\\ \nn &&
	+\frac{8}{135} \left(639 \phi ^4-618 \pi ^2 \phi
   ^2+47 \pi ^4\right) \left(\pi ^2-\phi ^2\right)^2
	\left[\text{Li}_3(e^{-2 i \phi
   })+\text{Li}_3(e^{2 i \phi })\right]
		\\ \nn &&
	+\frac{64}{135} i \phi  \left(22 \phi ^2-15 \pi
   ^2\right) \left(\pi ^2-\phi ^2\right)^3
	\left[\text{Li}_2(e^{2 i \phi
   })-\text{Li}_2(e^{-2 i \phi })\right]
		\\ \nn &&
+\frac{1168}{135} \phi ^2 \left(\pi ^2-\phi ^2\right)^4 \[\log (1-e^{2 i \phi })+\log (1-e^{-2 i \phi })\]
		\\ \nn &&
+\frac{752 \zeta (3) \phi ^8}{15}-\frac{16}{135} \left(970 \pi ^2 \zeta (3)+2493 \zeta (5)\right) \phi
   ^6
	\\ \nn &&
	+\frac{16}{45} \left(208 \pi ^4 \zeta (3)+1130 \pi ^2 \zeta (5)+5175 \zeta (7)\right) \phi
   ^4
	\\ \nn &&
	-\frac{16}{9} \left(2 \pi ^6 \zeta (3)+27 \pi ^4 \zeta (5)+414 \pi ^2 \zeta (7)+6615 \zeta (9)\right)
   \phi ^2
	\\ \nn &&
	-\frac{8}{135} \left(94 \pi ^8 \zeta (3)+888 \pi ^6 \zeta (5)+8190 \pi ^4 \zeta (7)+68040 \pi ^2
   \zeta (9)-1403325 \zeta (11)\right)
\eeqa

\section{Generalized $\eta$-functions}
\label{sec:eta}

We found that the solution of the QSC for arbitrary angles at weak coupling involves the following generalized $\eta$ functions
\beq
\label{defeta}
\eta_{s_1,\dots,s_k}^{z_1,\dots,z_k}(u)\equiv \sum_{n_1 > n_2 > \dots > n_k \geq 0}\frac{z_1^{n_1}\dots z_k^{n_k}}{(u+i n_1)^{s_1}\dots (u+i n_k)^{s_k}}
\eeq
which are a generalization of the multiple polylogarithms
\beq
{\rm Li}_{(s_1,\dots,s_k)}(z_1,\dots,z_k)=\sum_{n_1 > n_2 > \dots > n_k \geq 1}\frac{z_1^{n_1}\dots z_k^{n_k}}{n_1^{s_1}\dots n_k^{s_k}}
\eeq
For the case when all twists $z_i$ are set to 1, the $\eta$-functions were encountered in the weak coupling computations of \cite{Leurent:2013mr,Marboe:2014gma}. In our calculation of $\Gc$ we had to deal with the case where twists are present. Below we summarize some useful relations analogous to those found in \cite{Leurent:2013mr,Marboe:2014gma}.

Let us denote a solution of the equation
\beq
	f(u+i)-f(u)=h(u)
\eeq
as
\beq
	f=\Sigma(h)
\eeq
	
A useful property is
\beqa
\eta_{A,a}^{Z,z}=Z z(\eta_{A,a}^{Z,z})^{[2]}+
Z\frac{(\eta_{A}^{Z})^{[2]}}{u^a}
\eeqa
where $A$ is a set of indices $A_i$ and $Z$ in the superscript is a set of twists $Z_i$, while $z$ is a single complex number. The prefactor $Z$ in the r.h.s. denotes the product $\prod_i Z_i$. Using this relation we find
\beqa
\Sigma\(\frac{z^{-i u}}{(u+i n)^s}\)&=&-z^{-i u}\eta^z_s(u+ i n)\\
\Sigma\(\frac{z^{-i u}\eta_S^{Z}(u+i n+i)}{(u+i n)^s}\)&=&
-\frac{z^{-i u}}{Z}
\eta_{Ss}^{Z(z/Z)}(u+i n),\\
\Sigma
\[
v^{-i u}u^a \eta^{Zz}_{Ab}(u+i n)
\]
&=&
\Sigma\[
\(\frac{v}{zZ}\)^{-i u}u^a
\]
(zZ)^{-i u}
\eta^{Zz}_{Ab}(u+i n)\\
\nn&+&\Sigma
\[
\Sigma
\[
\(\frac{v}{zZ}\)^{-i u}u^a
\]^{[2]}
(zZ)^{-i u}
Z\frac{\eta^Z_A(u+i n+i)}{(u+i n)^b}
\]
\eeqa
In these expressions $a,s=1,2,3,\dots$ while $n$ is arbitrary.

Finally we have the 'stuffle' relations which express a product of two $\eta$ functions as a linear combination of some other $\eta$'s. They are obtained by splitting the region of summation  in the product of  $\eta$ functions and are directly analogous to those for polylogarithms or mutiple zeta values (see e.g. the pedagogical review \cite{Duhr:2014woa} and references therein):
\beqa
\eta_{\underline{s}}^{\underline{z}}\;\eta_{\underline{s}'}^{\underline{z}'}=\sum_{\underline{s}''} \eta_{\underline{s}''}^{\underline{z}''}
\eeqa
where in case two of the $s$ indices are combined in the r.h.s. the corresponding twists are mutiplied, exactly as in the stuffle relations for polylogarithms. For example,
\beq
	\eta_2^w \eta_3^z=\eta_5^{wz}+\eta_{2,3}^{w,z}+\eta_{3,2}^{z,w}
\eeq
The operations described above are essential for the iterative procedure of \cite{Gromov:2015vua} and should allow to run it to very high orders in the weak coupling expansion with any $\phi,\theta$.

\end{document}